\documentclass[10pt,letterpaper]{article}
\usepackage[margin=1in,nohead,centering]{geometry}
\usepackage{graphicx,color,wrapfig}
\usepackage{amssymb}
\usepackage{amsmath}
\usepackage{url}
\usepackage{xspace}
\usepackage{float}
\usepackage{subfigure}
\usepackage{graphics}

\newcommand{\aseq}{{\bf a}\xspace}

\newtheorem{example}{Example}
\newtheorem{definition}[example]{Definition}

\chardef\other=12
\def\mdeactivate{%
\catcode`\&=\other   \catcode`\#=\other
\catcode`\%=\other   \catcode`\~=\other
}
\def\mmakeactive#1{\catcode`#1=\active\ignorespaces}

{% The group delimits the text over which ^^M is active.
\mmakeactive\
\gdef\obeywhitespace{%
  \mmakeactive\^^M %
  \let^^M=\NewLine %
  \aftergroup\removebox %
  \obeyspaces %
}}

\def\NewLine{\par\indent}
\def\removebox{\setbox0=\lastbox}

\def\mverbatim{\par\begingroup\parindent=0em\tt\mdeactivate\obeywhitespace
\catcode`\|=0  %
}

\def\|{|}
\title{Expected degree for RNA secondary structure networks}
\author{P. Clote}
\date{}

\begin{document}

\title{Expected degree of RNA secondary structure networks}
\author{Peter Clote}
\date{}

\maketitle

\begin{abstract}
Consider the network of all secondary structures of a given RNA sequence,
where nodes are connected when the corresponding structures have base
pair distance one.
The expected degree of the network is the average number of
neighbors, where average
may be computed with respect to the either the uniform  or Boltzmann
probability.  Here we describe the first
algorithm, {\tt RNAexpNumNbors}, that can compute the expected number
of neighbors,  or expected network degree, of an input sequence. 
For RNA sequences from the
Rfam database, the expected degree is significantly less than the
CMFE structure, defined to have 
minimum free energy over all structures consistent with the
Rfam consensus structure.  The expected degree
of structural RNAs, such as purine riboswitches, paradoxically
appears to be smaller than that of random RNA, yet the difference
between the degree of the MFE structure and the expected degree
is larger than that of random RNA. Expected degree does not
seem to correlate with standard structural diversity measures of RNA, 
such as positional entropy, ensemble defect,  etc.
The program {\tt RNAexpNumNbors} is written in C, runs in cubic time
and quadratic space, and is publicly available at
\url{http://bioinformatics.bc.edu/clotelab/RNAexpNumNbors}.
\end{abstract}

\section{Introduction}
\label{section:intro}

%Nice math overview of small-world networks: 
%http://www.scholarpedia.org/article/Small-world_network 

Examples of small-world phenomena abound in the physical sciences.
In \cite{Watts.n98}, the neural connections of {\em C. elegans} were shown
to have both small mean path length between nodes and to have large
clique-like clusters, both hallmarks of small-world networks.
In \cite{VanNoort.er04}, the gene co-expression network of
{\em S. cerevisiae} was shown to have a scale-free, small-world 
architecture, where genes are linked in the network if they are co-expressed.
The network of metabolic pathways \cite{Wagner.pbs01,Ravasz.s02} and
of folding kinetics for the protein villin \cite{Bowman.pnas10} were
shown as well to have small-world properties, which collectively suggests
a topological robustness that appears to be intrinsic in biological 
processes.

In \cite{Wuchty.nar03}, Wuchty showed that the network of
secondary structures of {\em E. coli} phe-tRNA, with 
accession number RF6280 \cite{sprinzl}, is small-world,
by using the program {\tt RNAsubopt} \cite{wuchtyFontanaHofackerSchuster}
to analyze the ensemble of low energy secondary structures 
of phe-tRNA, having free energy approximately within 8.5 kcal/mol
(14 kT) of the minimum free energy. In this network, two nodes $s,t$
(secondary structures) are linked by an edge if $t$ can be obtained
from $s$ by removing or adding a single base pair, or obtained 
by a {\em shift move}.
A shift move, depicted in Figure~1 of \cite{Wuchty.nar03}, 
allows one to move one end of a 
base pair without moving the other.
%; i.e. $(i,j) \rightarrow (i,k)$ or $(i,j) \rightarrow (k,j)$.

Small-world networks satisfy a {\em connectivity} property, where the 
average shortest path distance (geodesic distance) is small relative 
to the network size -- typically logarithmic in network size. 
In the case of RNA
secondary structures, this property trivially holds, since the
number of secondary structures is generally exponential in the 
RNA sequence size $n$
\cite{steinWaterman}, while the base pair distance between
any two structures (and so path length) is at most $n$.

Wuchty showed as well that cluster sizes of the low energy ensemble of
secondary structures of phe-tRNA are larger than that of
(Erd\"os-Renyi) random graphs, and that
{\em clustering coefficient} $C(\nu)$ for phe-tRNA is inversely proportional
to node degree $k_{\nu}$ of node $\nu$; 
i.e. $C(\nu) \sim \frac{1}{k_{\nu}}$.
Here, for network node $\nu$, the
{\em clustering coefficient} $C(\nu)$ is defined as the fraction of
pairs of neighbors of $\nu$ which are connected by a network edge; i.e.
%\begin{eqnarray}
%\label{eqn:clusteringCoeff}
$C(\nu)  = \frac{2 \cdot n}{k_{\nu} \cdot (k_{\nu}+1)}$,
%\end{eqnarray}
where $n$ is the number of pairs of neighbors of $\nu$ that are connected by
a network edge.

Motivated originally by issues concerning the kinetics of RNA folding,
in this paper we study the {\em average} network degree, i.e. the expected
number of neighbors of a secondary structure for a given RNA sequence.
In contrast to the work of Wuchty \cite{Wuchty.nar03}, we consider a 
neighbor of
the secondary structure $s$ to be any structure obtained by
adding or deleting a single base pair from $s$, but {\em not} 
obtained by a {\em shift move}.
We describe the first algorithm, {\tt RNAexpNumNbors},
that can compute the
expected number of neighbors for a given RNA sequence, 
where expectation can be taken either with respect to the uniform
or the Boltzmann probability. 
Using dynamic programming, the C program {\tt RNAexpNumNbors} runs
in cubic time and quadratic space with respect to input sequence length.

The plan of the paper is as follows. To make basic notions absolutely clear,
Section~\ref{section:preliminaries} presents an illustrative example
of how to manually compute the expected network degree for a toy 8 nt
RNA sequence. Section~\ref{section:results} shows the inadequacy of this
exhaustive method, hence the need for an efficient program such as
{\tt RNAexpNumNbors}. In this section, we apply {\tt RNAexpNumNbors}
to nine noncoding RNA
families from the Rfam database \cite{Gardner.nar11}, and determine the
Pearson correlation between expected network degree and various RNA
structural measures. Definitions for these measures are given in the
Appendix, and the correlations are displayed in 
Tables~\ref{table:RfamExpNumNborsNineFamilies} 
to \ref{table:expLessMfeLessCmfe}.
Section~\ref{section:discussion} summarizes our main contributions and
poses some open questions. The recursions for our algorithm, 
{\tt RNAexpNumNbors}, are described in Section~\ref{section:methods},
and the detailed derivation for these recursions are given in the
Appendix.

\section{Preliminaries}
\label{section:preliminaries}

A secondary structure for an RNA nucleotide sequence
$\aseq = a_1,\dots,a_n$ is a set $s$ of Watson-Crick or wobble
base pairs $(i,j)$, containing neither base triples nor pseudoknots.
The number of base pairs in $s$ is denoted by $|s|$.
A secondary structure $s$ is {\em compatible}
with $\aseq$ if for every base pair $(i,j)$ in $s$, the pair
$(a_i,a_j)$ is contained in the set
of six canonical (Watson-Crick and wobble) base pairs. 
Throughout this paper, by {\em structure}, we mean a
secondary structure which is compatible with an arbitrary, but fixed
RNA sequence $\aseq$.

If $s,t$ are secondary structures of $\aseq$, then the {\em base pair distance},
$d_{BP}(s,t)$, is defined as $|s-t|+|t-s|$, i.e. the number of 
base pairs belonging to $s$ but not $t$, or vice versa.
Structures $s,t$ are said to be {\em neighbors}, if their base pair distance
is $1$, and to be $k$-{\em neighbors} if $d_{BP}(S,T) = k$.

Given an RNA sequence $\aseq = a_1,\ldots,a_n$,
the expected number $\langle N(\aseq)\rangle$
of neighbors for $\aseq$ is defined by
\begin{eqnarray}
\label{eqn:defExpNumNbors}
\langle N(\aseq) \rangle = \sum_{s} P(s) \cdot N(s) =  
\sum_s \frac{\exp(-E(s)/RT)}{Z} \cdot N(s)
\end{eqnarray}
where the sum is taken over all secondary structures $s$ of the input
sequence $\aseq$, where $P(s)$ is the probability of structure $s$, $N(s)$ is
the number of neighbors of $s$, $E(s)$ is the energy of $s$, 
$R= 0.001987$ kcal K$^{-1}$ mol$^{-1}$
and $T = 310.15$ K. At times we may write $N_s$
in place of $N(s)$, especially in Section~\ref{section:methods} 
and the Appendix.
Often the RNA sequence $\aseq$
is clear from the context and so omitted, and we may correspondingly
designate the expected number of neighbors by $\langle N \rangle$.

In this paper, we describe novel, efficient (cubic time, quadratic
space)  algorithms to compute
the expected number of neighbors, with respect to three energy models.
Model A assigns an energy of 0 to each structure. It follows that
the partition function $Z = \sum_{s} \exp(-E(s)/RT)$ is simply the
number of structures, and so the probability
$P(s) = \frac{\exp(-E(s)/RT)}{Z}$ is the {\em uniform} probability. 
Model B, first proposed by Nussinov and Jacobson \cite{nussinovJacobson},
assigns an energy of $-1$ to each base pair, so the energy
of a structure having $k$ base pairs is $-k$. Model C, commonly known
as the Turner energy model \cite{turner,xia:RNA}, assigns negative,
stabilizing free energies
to {\em stacked} base pairs, and positive, destabilizing free energies to
hairpin loops, bulges, internal loops and multiloops. The energy parameters
are derived from UV absorption (optical
melting) experiments,  except for a multiloop affine energy approximation.
In our current software {\tt RNAexpNumNbors}, 
we employ the Turner 1999 energy parameters \cite{turner,xia:RNA,Turner.nar10}
{\em without} dangles. Accounting for dangles, or single-stranded, stacked 
nucleotides, would add considerable complexity to our dynamic programming
algorithm; indeed, at 
present, it is unclear how this might even be done.  

\begin{figure*}
\centering
\begin{minipage}{0.45\textwidth}
\mverbatim
ACGUACGU
(......)   4.40 kcal/mol
((....))   1.90 kcal/mol
..(....)   5.70 kcal/mol
.(....).   3.60 kcal/mol
........   0.00 kcal/mol
|mendverbatim
\end{minipage}
\hskip 1cm
\begin{minipage}{0.45\textwidth}
\fbox{
\includegraphics[width=0.4\textwidth]{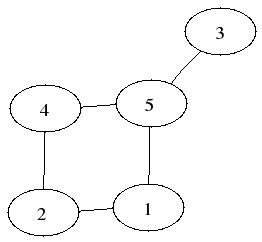}
}%end{fbox}
\end{minipage}
\caption{{\em (Left)} All possible secondary structures of the 8-mer
ACGUACGU with corresponding free energies in kcal/mol, computed by
{\tt RNAeval} from the Vienna RNA Package \cite{hofacker:ViennaWebServer} using
the Turner 1999 energy model without dangles \cite{turner,xia:RNA}.
{\em (Right)} Graph representation of neighborhood network, where
nodes 1,2,3,4,5 respectively represent the secondary structures
{\tt (......)}; {\tt ((....))}; {\tt ..(....)};  {\tt .(....).};
{\tt ........}. The {\em expected number of neighbors} corresponds to
the expected degree in the graph with respect to the uniform probability.
In computing the expectation with respect to the Boltzmann probability,
each node is weighted by its Boltzmann factor divided by the partition
function.
}
\label{fig:toyComputationExpNumNborsAndNeatoGraph}
\end{figure*}

Although the Turner energy model is the only physically realistic model,
involving enthalpic and entropic considerations, our dynamic programming
method is complicated and more easily explained by first addressing model
A, then B, then C. 
To illustrate the definition of expected number of neighbors, given
in equation (\ref{eqn:defExpNumNbors}), consider the 8 nt RNA sequence
ACGUACGU, all of whose secondary structures can be generated
by the program {\tt RNAsubopt} \cite{wuchtyFontanaHofackerSchuster} --
see Figure~\ref{fig:toyComputationExpNumNborsAndNeatoGraph}.

%{%
%\setlength{\fboxsep}{0pt}%
%\setlength{\fboxrule}{1pt}%
%\fbox{\includegraphics[options]{image}}%
%}%

Due to steric constraints, by definition each hairpin loop is required to 
have at least three unpaired bases. It follows that the first structure
in Figure~\ref{fig:toyComputationExpNumNborsAndNeatoGraph}
has 2 neighbors, i.e. either the empty structure, obtained by removing
base pair $(1,8)$, or the second structure, obtained by adding the base
pair$(2,7)$. The second structure has 2 neighbors, obtained by removing
base pair $(1,8)$ or $(2,7)$. 
The third structure has only 
one neighbor, obtained by removing base pair $(3,8)$, while the 
fourth structure has two neighbors, 
obtained by either removing base pair $(2,7)$ or adding base pair $(1,8)$.
In contrast the fifth structure, which is
empty, has 3 neighbors, obtained by adding either base pair $(1,8)$,
$(2,7)$ or $(3,8)$. If the energy of each structure is $0$, as in Model A,
then equation (\ref{eqn:defExpNumNbors}) yields
$\frac{2+2+1+2+3}{5} = 2$; i.e. 
the {\em uniform} expected number of neighbors is $2$.

In Model B, usually called the Nussinov energy model, the partition
function $Z$ is the sum of the Boltzmann factors $\frac{\exp(-E(s)/RT}{Z}$,
hence
\begin{eqnarray*}
Z &=& \exp(1/RT) + \exp(2/RT) + \exp(1/RT) + \\
&&  \exp(1/RT) + \exp(0/RT) = 41.8611.
\end{eqnarray*}
Thus, with respect to the Nussinov energy model,
the {\em Boltzmann} expected number of neighbors is 
\begin{eqnarray*}
\langle N \rangle &=& \frac{\exp(1/RT)}{Z} \cdot 2 +
\frac{\exp(2/RT)}{Z} \cdot 2 +
\frac{\exp(1/RT)}{Z} \cdot 1 + \\
&&\frac{\exp(2/RT)}{Z} \cdot 2 +
\frac{\exp(0/RT)}{Z} \cdot 3 = 
\frac{79.656} {41.861} = 1.903 .
\end{eqnarray*}

In the Turner energy model, the free energies
of the five structures are 4.4, 1.9, 5.7, 3.6 and 0.0 kcal/mol, reflecting the
fact that the minimum free energy structure for this toy 8-mer is the
empty structure. The partition function is
\begin{eqnarray*}
Z &=& \exp(-4.4/RT)+\exp(-1.9/RT)+\exp(-5.7/RT)+\\
  & &\exp(-3.6/RT)+\exp(-0/RT) =    1.049626 .
\end{eqnarray*}
Thus, with respect to the Turner energy model, the {\em Boltzmann}
expected number of neighbors is
\begin{eqnarray*}
\langle N \rangle &=& \frac{\exp(-4.4/RT) \cdot 2}{Z} +
\frac{\exp(-1.9/RT) \cdot 2}{Z} +
\frac{\exp(-5.7/RT) \cdot 1}{Z} + \\
&&\frac{\exp(-3.6/RT) \cdot 2}{Z} +
\frac{\exp(-0.0/RT) \cdot 3}{Z} =
\frac{ 3.0992}{1.0496} = 2.9526 .
\end{eqnarray*}

In the sequel, the phrase {\em Boltzmann expected number of neighbors}
will mean that the expected value is computed with respect to Boltzmann
probability using the Turner 1999 energy model without dangles -- i.e.
Model C. A future version of {\tt RNAexpNumNbors} will alternatively
support the Turner 2000 parameters.

\section{Results}
\label{section:results}

It is straightforward, to automate the previous manual computations, 
and thus determine the expected number of neighbors for a given RNA sequence by exhaustively
listing all secondary structures and their free energies with the
program {\tt RNAsubopt} \cite{wuchtyFontanaHofackerSchuster}. This approach is
only posssible for a sufficiently small RNA sequence, since the number of
secondary structures is exponential in the sequence length 
\cite{steinWaterman}.  Binning the output of {\tt RNAsubopt} according
to the number of neighbors of each secondary structure, we can determine
the relative frequency of structures of the 32 nt selenocysteine insertion
sequence (SECIS) element {\tt fruA} with sequence
CCUCGAGGGG AACCCGAAAG GGACCCGAGA GG and the 27 nt bistable switch
with sequence CUUAUGAGGG UACUCAUAAG AGUAUCC and two meta-stable structures
{\tt .......((((((((....))))))))} having -10.30 kcal/mol and
{\tt ((((((((....)))))))).......} having -9.90 kcal/mol.

\begin{figure*}
\centering
\includegraphics[width=0.45\textwidth]{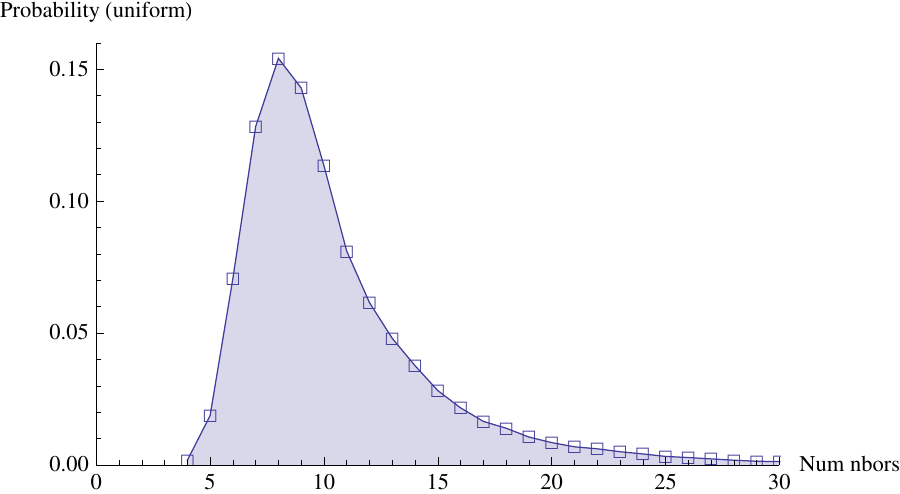}
\includegraphics[width=0.45\textwidth]{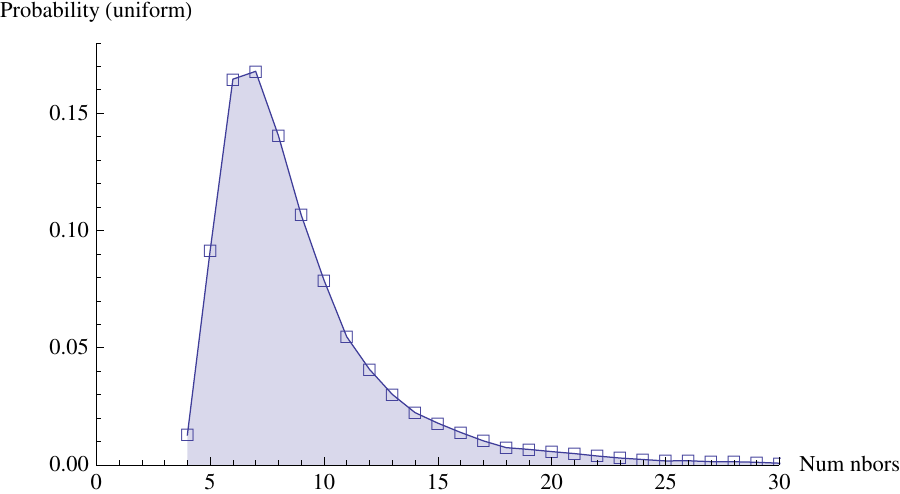}
\caption{ {\em (Left)} Relative frequency of the number of structures
for the  32 nt selenocysteine insertion sequence (SECIS) element
fruA, obtained by by exhaustive generation of
all secondary structures. The number of neighbors is given on the $x$-axis,
while the proportion of structures having a given number of neighbors is given
on the $y$-axis.
The expected number of neighbors with respect to the
uniform distribution is
10.657 with standard deviation 4.777, while that for the Boltzmann probability
is 10.001 with standard deviation 0.058. 
{\em (Right)} Similar analysis for a 27 nt bistable switch.
The expected number of neighbors with respect to the
uniform distribution is 9.14 with with standard deviation 4.56,
while that for the Boltzmann probability
is 11.92 with standard deviation 0.549.
}
\label{fig:FruAexpNumNbors}
\end{figure*}

Figure~\ref{fig:FruAexpNumNbors}
displays the relative frequency with respect to the uniform probability,
that structures for {\tt fruA} (left panel) resp. the bistable switch
(right panel) have a given number of neighbors. In particular, 
110,124 of the 971,299 secondary structures of {\tt fruA} 
($11.3\%$) have exactly 10 neighbors. The analogous computation with respect
to the Boltzmann probability indicates that $99.8\%$ of the structures 
have exactly 10 neighbors -- of course, this means that essentially all of
the low energy structures have 10 neighbors.
The expected number of neighbors for {\tt fruA} with respect to the
uniform distribution is
$10.657$ (stdev $4.777$), while that for the Boltzmann probability
is $10.001$ (stdev $ 0.058$).  
For the 27 nt bistable switch, 
30,609 of the 186,105 secondary structures 
($16.4\%$) have exactly 6 neighbors, with respect to the
uniform probability, while $88.36\%$ of the structures have
12 neighbors with respect to the Boltzmann probability.
The MFE structure having -10.30 kcal/mol has 12 neighbors,
while the meta-stable structure having -9.90 kcal/mol has 11 neighbors.

\begin{figure*}
\centering
\includegraphics[width=0.45\textwidth]{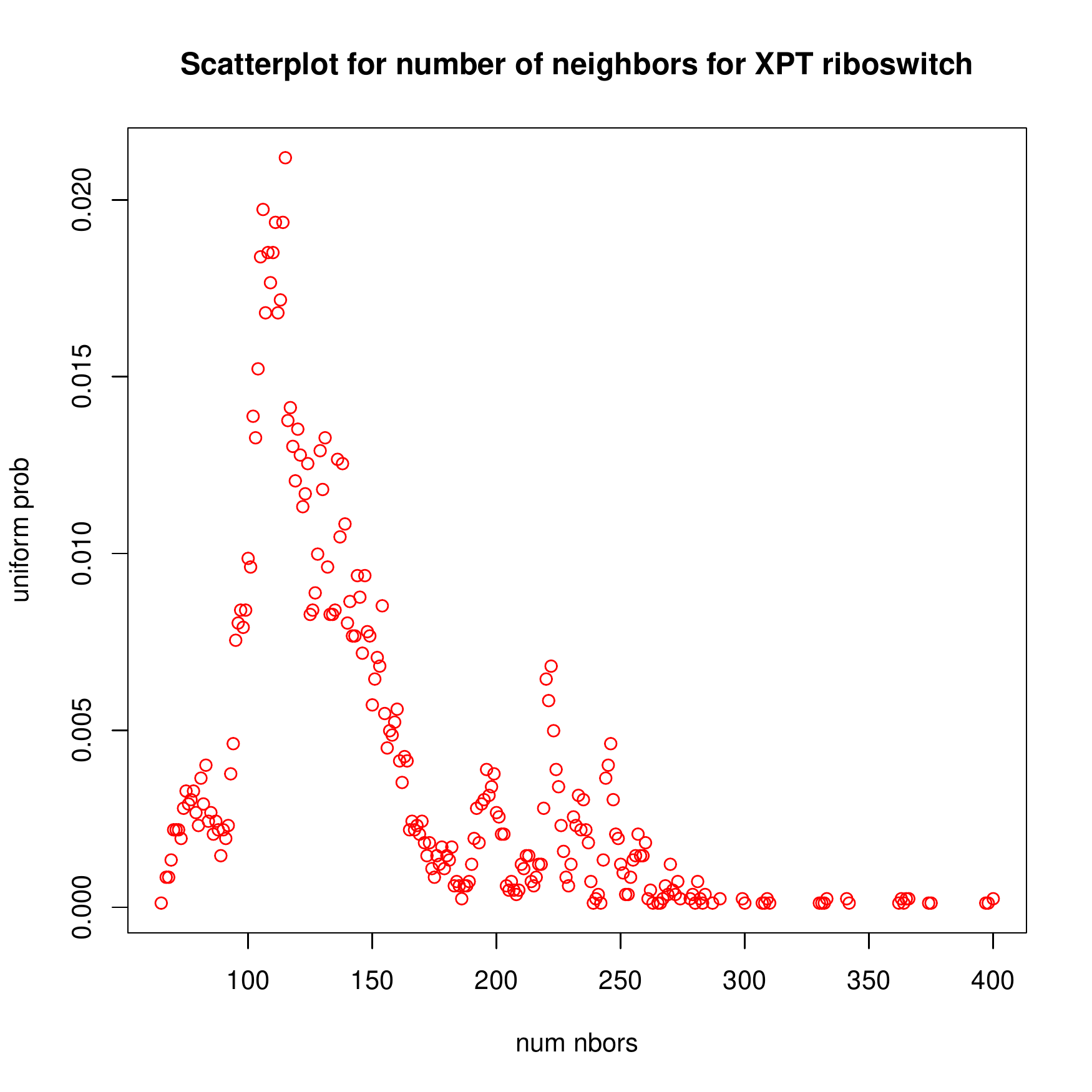}
\includegraphics[width=0.45\textwidth]{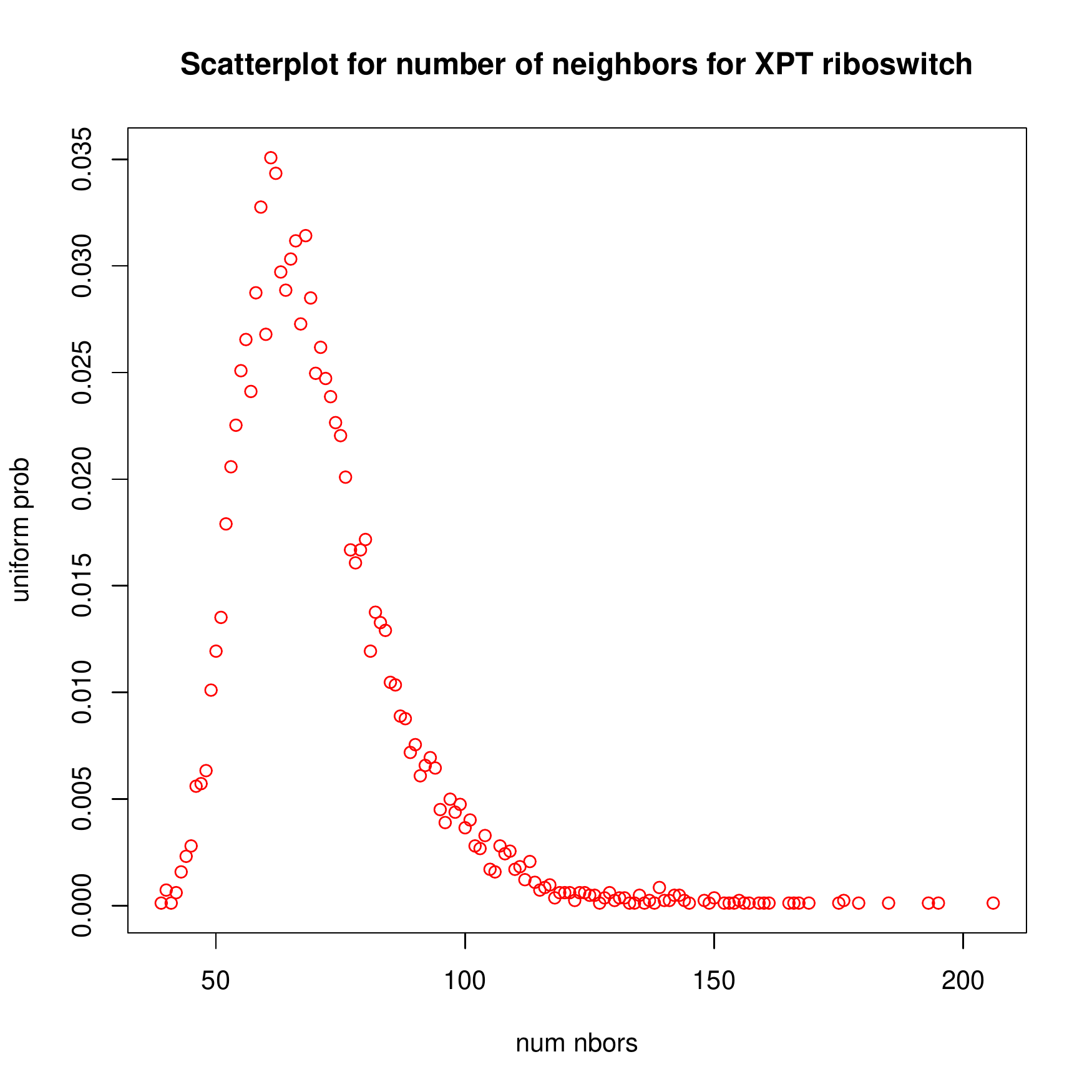}
\caption{ {\em (Left)} Relative frequency of the number of structures
for the 161 nt xanthine phosphoribosyltransferase (XPT) riboswitch, estimated
by using {\tt RNAsubopt} to sample 8212 structures having free energy within 
5 kcal/mol of the MFE. Structures were binned according to number of neighbors,
yielding a mean of $138.848$ with standard deviation of $45.457$, compared
with the exact value of
$153.179$ obtained with {\tt RNAexpNumNbors}.
%Each sampled structure received a weight proportional
%to its Boltzmann factor, resulting in
%an expected number of neighbors of
%$150.866 \pm 48.237$, compared with the exact value of
%$153.179$ obtained with {\tt RNAexpNumNbors}.
{(\em Right)} Analogous graph, where 8000 structures were uniformly sampled from
the ensemble of all structures for the XPT riboswitch. This data
yields an estimate of $70.326 \pm 16.249$ for the
uniform expected number of neighbors of XPT, compared with the exact
value of $61.040$ obtained with {\tt RNAexpNumNbors}.
}
\label{fig:sampledXPThistogram}
\end{figure*}

For larger sequences, one can use {\tt RNAsubopt} to sample those
structures, whose free energy lies within a user-specified bound
of that of the MFE structure.  For the 161 nt
xanthine phosphoribosyltransferase (XPT) riboswitch,
depicted in Figure 1 of \cite{Serganov.cb04}, we used {\tt RNAsubopt} to
sample 8212 structures having free energy within 5 kcal/mol of the MFE.
%Computing the Boltzmann expected number of neighbors from this set, and 
%normalizing by the sum of Boltzmann factors of the structures sampled, we obtain
%a mean of $150.866$ and standard deviation of $48.237$, compared with the
%exact value of $153.179$ obtained with {\tt RNAexpNumNbors}. 
By computing the relative frequency that sampled structures have exactly $k$
neighbors, we obtain the density plot given in the left panel of 
Figure~\ref{fig:sampledXPThistogram}, which yields the estimated
mean $138.848$ (stdev $45.457$), compared with the
correct value of $153.179$ obtained with {\tt RNAexpNumNbors}. 
By increasing the free energy bound of 5 kcal/mol to 10 kcal/mol, one could
obtain an improved graph and somewhat more accurate estimate of the mean;
however, this comes at a severe computational cost, since the number of 
structures grows exponentially in the free energy bound. (Compare the
left panel of
Figure~\ref{fig:sampledXPThistogram} with that of Figure~4 from
\cite{Wuchty.nar03}.)

We wrote a program in C to count the number of secondary structures for an input
RNA sequence and output a user-specified number of structures,
sampled with respect to the uniform distribution. In this fashion, we
generated 8000 structures from the ensemble of all structures of the XPT riboswitch,
and determined an estimate of $70.326$ with standard deviation of $16.249$ for the
uniform expected number of neighbors of XPT riboswitch. The 
exact value is $61.040$, as determined by {\tt RNAexpNumNbors}.
The right panel of Figure~\ref{fig:sampledXPThistogram} depicts the graph of
uniform distribution of the number of neighbors for XPT.

It follows from this illustrative example that
there is no current method, apart from the algorithm {\tt RNAexpNumNbors}
of this paper, which can accurately compute the expected number of
neighbors for a given RNA sequence. The dynamic programming recursions are
described in Section~\ref{section:methods}, while full details of the 
derivation of the recursions is given in the Appendix.

\subsection*{Analysis of selected Rfam families}

In this section, we apply {\tt RNAexpNumNbors} to compute the
expected number of neighbors, both with respect to the uniform and
Boltzmann probability, for noncoding RNA from the 
Rfam 11.0 database \cite{Gardner.nar11}. 
The nine Rfam families are
5S ribosomal RNA (RF00001), 
U2 spliceosomal RNA (RF00004),
transfer RNA (RF00005),
type III hammerhead ribozyme (RF00008),
Selenocysteine insertion sequence 1 (RF00031),
small nucleolar RNA (RF00045),
purine riboswitch (RF00167),
HIV primer binding site (RF00375),
molybdenum cofactor riboswitch (RF01055).

\begin{figure}
\centering
\includegraphics[width=0.32\textwidth]{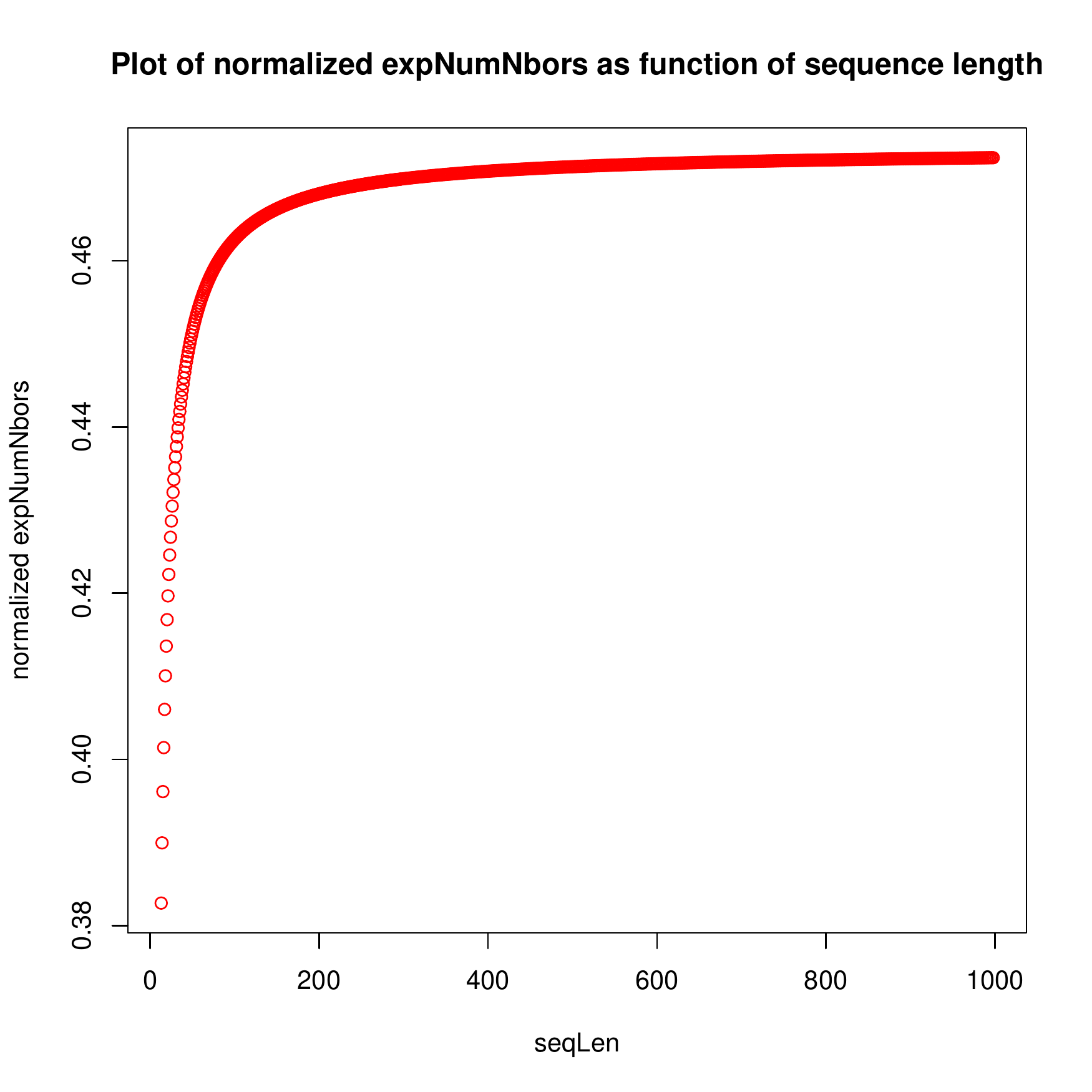}
\includegraphics[width=0.32\textwidth]{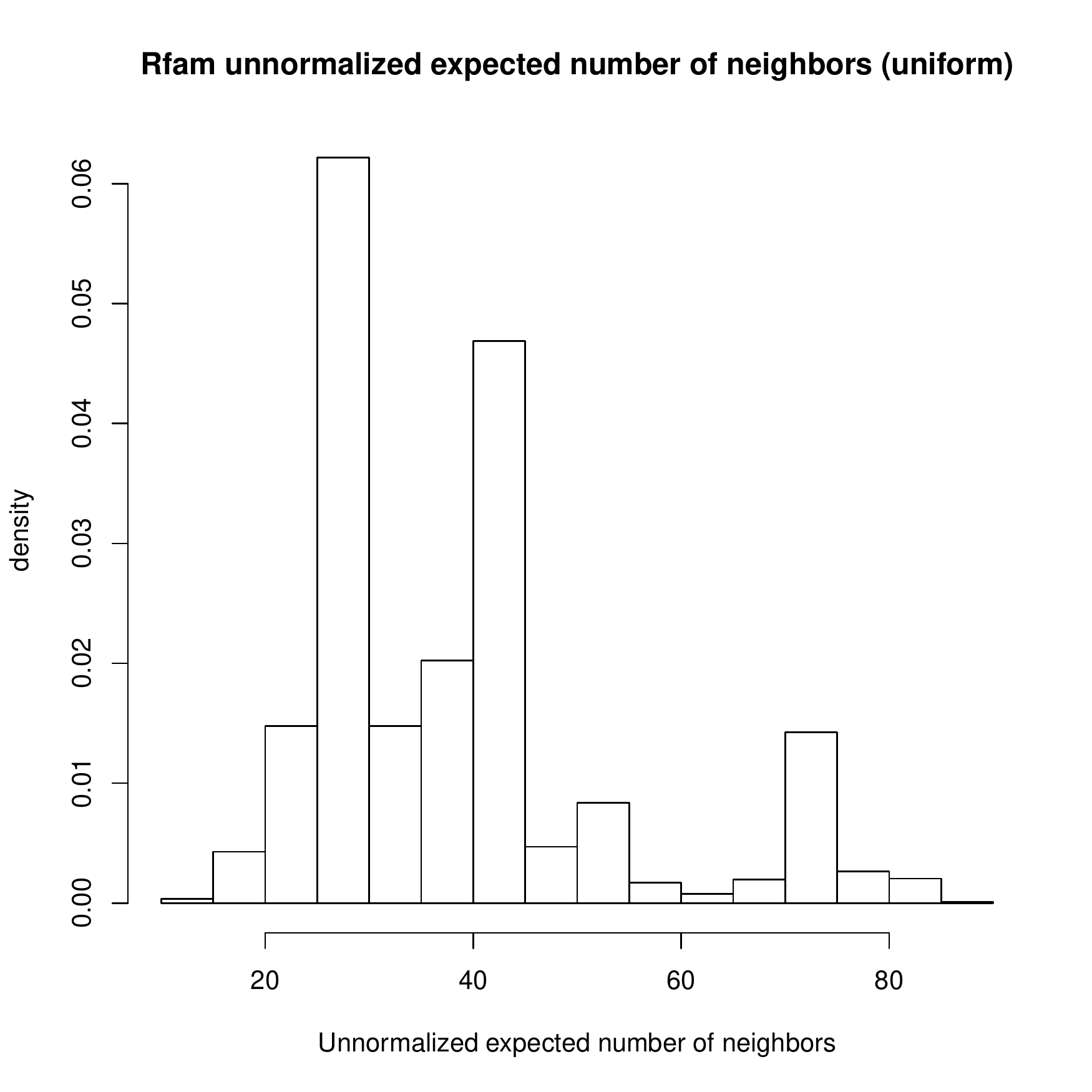}
\includegraphics[width=0.32\textwidth]{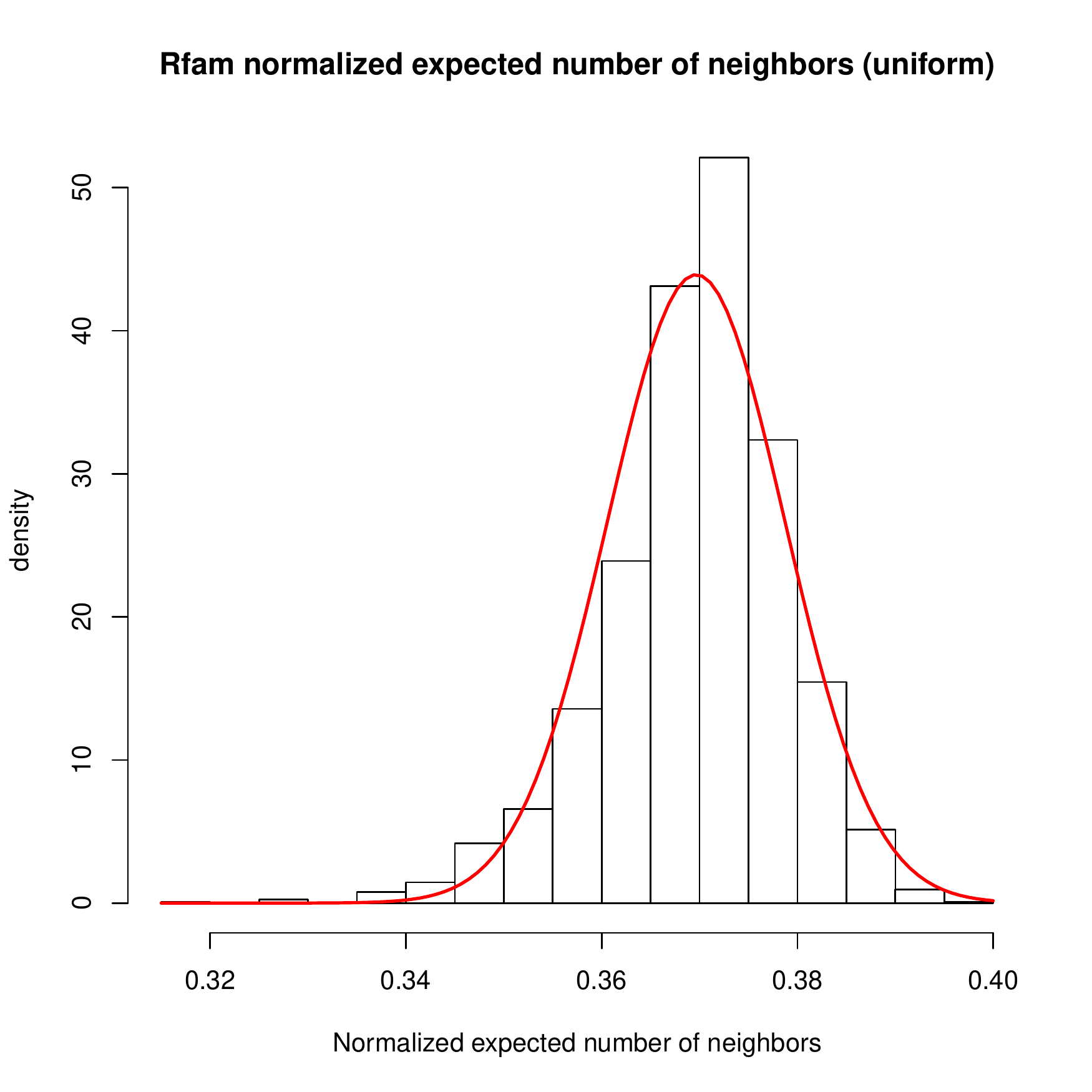}
\caption{{\em (Left)}
Plot of the normalized expected number of neighbors as a function
of sequence length, for homopolymers of length 1 to 1000 nt, obtained by
dividing the expected number of neighbors by sequence length.
Apparent asymptotic value is $\approx 0.4724$. 
{\em (Center)}
Relative frequency of the
expected number of neighbors pooled from all sequences in
seed alignments from nine Rfam families (see text).
{\em (Right)}
Relative frequency of the
{\em normalized} expected number of neighbors pooled from all sequences in
seed alignments from nine Rfam families, with a fitted normal distribution.
%Homopolymer homopolymer recursions
%Determine asymptotic value of $\frac{Q(n)}{n\cdot N(n)}$, where
%by equation (\ref{eqn:Qij_inductiveCase1}), we have
%$Q(n)=0$ for $-1 \leq n \geq 4$, and for $n>4$ we have
%$Q(n) =
%Q(n-1) + \sum_{k=0}^{n-6} 2 N(k) \cdot N(n-2-k) + 
%Q(k) \cdot N(n-2-k) + N(k) \cdot Q(n-2-k)$,
%and where by equation (\ref{eqn:Nij_inductiveCase1}), we have
%$N(n)=1$ for $-1 \leq n \geq 4$, and for $n>4$ we have 
%$N(n) = N(n-1) + \sum_{k=0}^{n-6}  N(k) \cdot N(n-2-k)$. Note that
%the asymptotic value of $N(n)$ is known from \cite{steinWaterman}.
}
\label{fig:asymptoticNormalizedExpNumNborsHomopolymer}
\end{figure}

Since it is clear that
longer RNA sequences in general have a larger number of neighbors,
these values should be normalized for comparative purposes. The left panel of
Figure~\ref{fig:asymptoticNormalizedExpNumNborsHomopolymer} depicts
the expected number of neighbors, {\em normalized} by dividing by
sequence length, for homopolymers of length  10 to 1000.  In this context,
a {\em homopolymer} is a sequence, where any two positions $i<j$ can
form a base pair, as long as $j-i \geq 4$, which ensures a minimum of
at least three unpaired bases in each hairpin loop. 
Figure~\ref{fig:asymptoticNormalizedExpNumNborsHomopolymer} clearly indicates
that the {\em normalized} expected number of
neighbors is asymptotically a constant in the homopolymer case, with
asymptotic value $\approx 0.4724$. The center [resp. right] panels of
Figure~\ref{fig:asymptoticNormalizedExpNumNborsHomopolymer} plot
the uniform [resp. {\em normalized} uniform]
expected number of neighbors pooled from all sequences in
the seed alignments of the nine Rfam families:
RF00001, RF00004, RF00005, RF00008, RF00031,
RF00045, RF00167, RF00375, RF01055. Sequence lengths in
this pooled set range from 40 nt to 225 nt, with a mean of
$104.21 \pm 39.96$. No visible pattern emerges in the center panel,
corresponding to {\em unnormalized} values. In
normalizing by dividing the expected number of neighbors
by sequence length, the values appear to be normally distributed
with a mean of $0.3697 \pm 0.0091$.
These arguments justify our normalized expected number of neighbors,
when comparing RNAs of different lengths from different Rfam families.

Table~\ref{table:RfamExpNumNborsNineFamilies} presents the averages,
taken over all sequences in the seed alignment of each of nine selected
families in the Rfam 11.0 database \cite{Gardner.nar11}, 
of the expected number of neighbors
with respect to both the uniform probability and to the Boltzmann
probability.  
Though small, the different values for the expected number of neighbors
are statistically significant. For instance, the
$p$-value is $2.00723 \cdot 10^{-46}$, for the
2-tailed T-test of equality for the average (uniform) normalized
number of neighbors for sequences from the seed alignment of  RF00001
(712 5S rRNA sequences) and RF00005 (960 tRNA sequences).

The normalized expected number of neighbors (Boltzmann probability)
appears to be completely uncorrelated with 
the normalized expected number of neighbors (uniform probability) --
taken over the pooled data from nine Rfam families, the Pearson
correlation is only $0.028191$, as shown in 
Table~\ref{table:normNumNbors}. This table considers as well
the number of neighbors of the minimum free energy
(MFE) structure and the the {\em constrained} minimum free energy
(CMFE), where for the latter, we used {\tt RNAfold -C}
from the Vienna RNA Package \cite{hofacker:ViennaWebServer} to compute
the structure having minimum free energy among all structures that
are {\em compatible} with the Rfam consensus structure. By this, we
mean that the CMFE structure $s$ obtained by {\tt RNAfold -C} does
not {\em conflict} with the constraints; if position $k$ is constrained
to be unpaired, then position $k$ must be unpaired in $s$, and if
$(x,y)$ is constrained to be a base pair, then for every base pair
$(i,j)\in s$, we have that if $\{ i,j\}$ and  $\{ x,y\}$ have non-empty
intersection, then $i=x$ and $j=y$, and that it is not the case that
$x<i<y<j$ or $i<x<j<y$. Note that {\tt RNAfold -C} does {\em not} require that
the base pair $(x,y)$ from the constraint belong to $s$, but only that
$s$ not conflict with the constrained base pair.

As far as we can determine, the expected number of neighbors,
equivalent to network degree,  seems to be
orthologous to other measures. In particular, there appears to be no relation 
between length-normalized
{\em Boltzmann } expected number of neighbors  (EXPB),
length-normalized 
{\em uniform } expected number of neighbors  (EXPU), GC-content,
minimum free energy, sequence length, 
positional entropy \cite{Huynen.jmb97},
expected number of base pairs \cite{Waldispuhl.jcb07},
ensemble defect \cite{Dirks.nar04},
expected base pair distance \cite{GarciaMartin12}, 
etc. Table~\ref{table:noCorrWithOtherMeasures} shows the absence
of correlation between EXPB and various structural diversity
measures (data for other measures not shown). See the Appendix for
definitions of positional entropy, expected number of base pairs, 
ensemble defect, expected base pair distance.

Table~\ref{table:corrNormDiffA} presents correlations between
the expected number of neighbors and other measures, defined as follows.
E: minimum free energy;
MFE-EXPB: (number of neighbors of MFE structure minus the Boltzmann expected number of neighbors)
divided by sequence length -- i.e. length-normalized;
EXPB: Boltzmann expected number of neighbors, divided by sequence length;
MFE-EXPU: (number of neighbors of MFE structure minus the uniform expected number of neighbors)
divided by sequence length;
MFE: number of neighbors of the MFE structure divided by sequence length;
SeqLen: sequence length. 

Some correlations are obvious; 
e.g. corr(E,SeqLen) = -0.8621 indicates that as sequence length
increases, the minimum free energy decreases.  The correlation of 
0.8724 between MFE-EXPB and MFE-EXPU is significant and surprising, 
since there is essentially no
correlation between EXPB and EXPU, as shown in
Table~\ref{table:normNumNbors}.
The positive correlation of 0.8721 between MFE-EXPB and MFE, 
and of 0.999436 between MFE-EXPU and MFE seems surprising.
However, since EXPB [resp. EXPU] values of members of a given Rfam family
appear to be close to the family 
average (see Table~\ref{table:RfamExpNumNborsNineFamilies}),
taken together this
suggests that the value of MFE essentially defines the values
of MFE-EXPB [resp. MFE-EXPU]. Finally, the correlation between
MFE and EXPB is likely due to the fact that MFE-EXPB is small,
in general, as shown in Table~\ref{table:expLessMfeLessCmfe}. 

Finally, Table~\ref{table:expLessMfeLessCmfe} compares the 
number of neighbors of the MFE and CMFE structures with the
expected number. In this table, 
MFE [resp. CMFE] stands for
the {\em length-normalized} number of neighbors for the minimum free
energy structure [resp. the structure having minimum free energy among
those structures that are consistent with the Rfam consensus structure].
EXP stands for the {\em length-normalized} expected number of neighbors,
as computed by {\tt RNAexpNumNbors}, and BPdist is the {\em length-normalized}
base pair distance between the MFE structure and the CMFE structure.
The table shows perhaps surprisingly that the MFE structure does {\em not}
have significantly more neighbors than the Boltzmann expected number;
however, the CMFE structure does. The Rfam consensus structure is
often taken as the gold standard in RNA benchmarking studies; however,
since Rfam base pairs are inferred only by covariation found in a multiple
alignment, we take the CMFE structure as representative of the native
structure.  It appears 
significant that the CMFE structure has significantly more neighbors than
the MFE structure. Another striking observation is large value of
EXPB-MFE for the two riboswitch families present in the collection
of Rfam sequences we investigated --
purine riboswitch (RF00167) and molybdenum cofactor riboswitch (RF01055).
If this finding holds up under careful scrutiny of all riboswitch families
in Rfam, then perhaps {\tt RNAexpNumNbors} could be used as a 
tool, along with {\tt RNAbor} \cite{FreyhultMoultonClote:RNAbor}
and {\tt FFTbor} \cite{FFTbor} to detect conformational switches.

\subsection*{Z-scores}

\begin{figure*}
\centering
\includegraphics[width=0.3\textwidth]{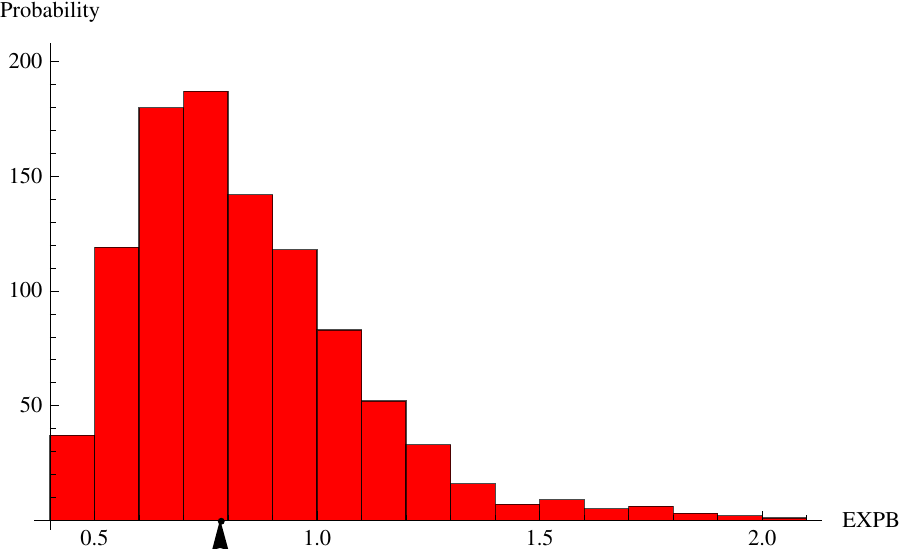}
\includegraphics[width=0.3\textwidth]{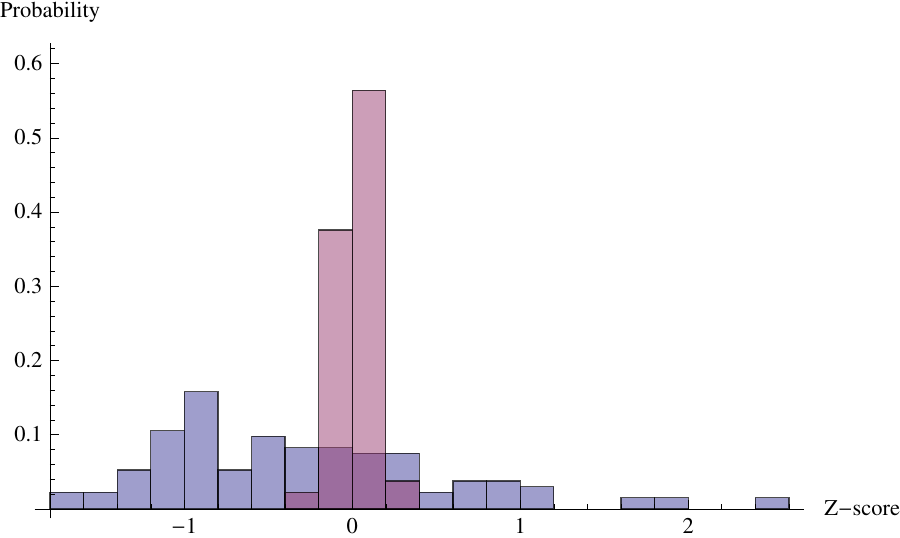}
\includegraphics[width=0.3\textwidth]{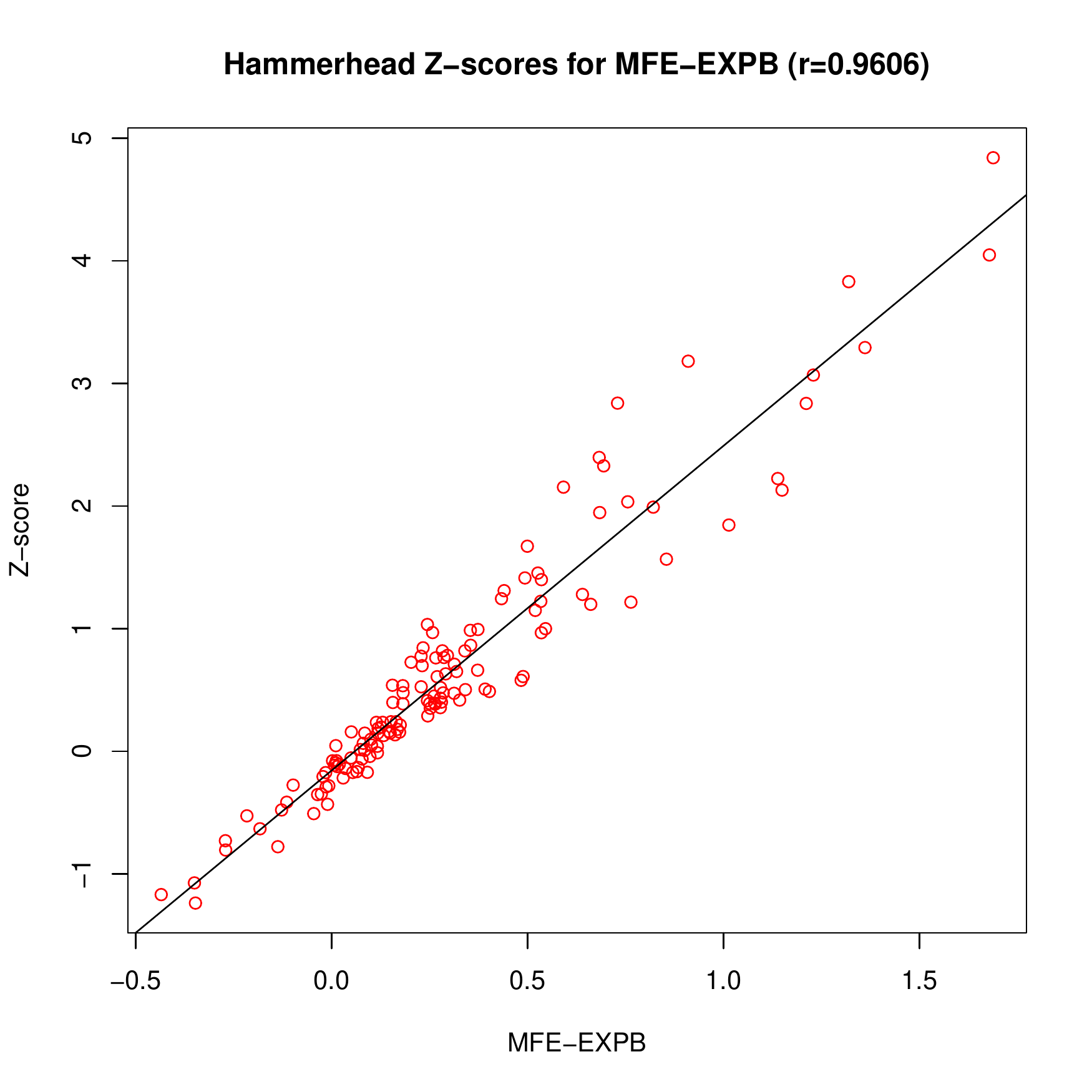}
\caption{{\em (Left)} Distribution of length-normalized Boltzmann
expected number of neighbors (EXPB) values for 1000 random
RNAs, where the arrow head marks the EXPB value wild type
purine riboswitch with EMBL accession code AE005176.1/1159509-1159606. 
Random RNA sequences were generated to have the same dinucleotides
as that of the wild type purine riboswitch by using the
Altschul-Erikson algorithm \cite{altschulErikson:dinucleotideShuffle}. 
{\em (Center)} Histogram of Z-scores, given by the
formula  (`wild type EXPB' - `mean EXPB of
random RNA') divided by `stdev EXPB of random RNA', 
for the {\em length-normalized}
expected number of neighbors for the 133 purine riboswitches from
the seed alignment of Rfam family RF00167.
For each riboswitch, its expected number $x$ of neighbors was computed,
as well as the mean $\mu$ and standard deviation $\sigma$ of the 
expected number of neighbors for 100 random RNA sequences having the
same dinucleotides, obtained by the Altschul-Erikson algorithm.
Z-scores for expectations
with respect to the Boltzmann [resp. uniform] probability appear in
blue [resp. red], with an overall mean  of $-0.291$ [resp.  $+0.019$].
{\em (Right)} Scatter plot of Z-scores for MFE-EXPB, where for a given
RNA sequence $\aseq$, MFE-EXPB  is defined as the length-normalized
number of neighbors of the MFE structure minus the length-normalized
expected number of neighbors.
For each riboswitch, MFE-EXPB was
computed, as well as that for 100 random RNA sequences having the
same dinucleotides, obtained by the Altschul-Erikson algorithm.
Corresponding Z-scores
$z=\frac{x-\mu}{\sigma}$ were computed, where $x$ is MFE-EXPB and
$\mu, \sigma$ are respectively the mean, standard deviation of 
corresponding MFE-EXPB values for the randomized RNAs. 
Pearson correlation is 0.9606.
}
\label{fig:Zscores}
\end{figure*}

The left panel of Figure~\ref{fig:Zscores} depicts the relative frequency
for the EXPB value for 1000 random
RNAs,  where EXPB denotes the length-normalized Boltzmann
expected number of neighbors.
The arrow head in the graph marks the EXPB value for wild type
purine riboswitch with EMBL accession code AE005176.1/1159509-1159606. 
Random RNA sequences were generated to have the same dinucleotides
as that of the wild type purine riboswitch by using the
Altschul-Erikson algorithm \cite{altschulErikson:dinucleotideShuffle}. 
Wild type purine riboswitch with EMBL accession code 
AE005176.1/1159509-1159606 has EXPB value 0.782112, while the average
EXPB of the 1000 randomized RNAs is $0.837026$ with standard deviation
$0.253572$. It follows that the Z-score for this riboswitch is
$z = \frac{0.782112-0.837026}{0.253572} = -0.21656$ -- i.e. most random
RNAs have larger EXPB values than this riboswitch. This situation is
in fact typical, as shown by the center panel of 
Figure~\ref{fig:Zscores}, which depicts the relative frequency
of Z-scores for the length-normalized 
expected number of neighbors for 133 purine riboswitch
sequences from the seed alignment of Rfam family RF00167.
For each riboswitch sequence, the expected number $x$ of neighbors was 
computed, and well as the expeccted number $x_1,\ldots,x_{100}$ of
100 random RNA sequences having the
same dinucleotides, obtained by the Altschul-Erikson algorithm.
Z-scores were computed
as $z = \frac{x-\mu}{\sigma}$, where
$\mu$ is the mean $\mu$ and $\sigma$ is the standard deviation 
$x_1,\ldots,x_{100}$. Z-scores computed with respect to
the Boltzmann probability with overall mean $-0.291$
appear in blue, while those computed with respect to the uniform
probability with overall mean $+0.019$  appear in red. It follows
that purine riboswitches 
tend to have a {\em lower} expected number of neighbors
than do their randomizations.

The right panel of Figure~\ref{fig:Zscores} depicts the relation between
MFE-EXPB and its Z-scores, described as follows.
For each purine riboswitch $\aseq$, the {\em length-normalized} difference
MFE-EXPB between the number of neighbors of the MFE structure and the
Boltzmann expected number was computed, as well that for
100 random RNA sequences $\aseq_1,\ldots,\aseq_{100}$ having the
same dinucleotides, obtained by the Altschul-Erikson algorithm.
Let $x$ be the MFE-EXPB value for $\aseq)$, 
and let $\mu$ [resp. $\sigma$] denote the
mean [resp. standard deviation] for the MFE-EXPB values of
the random RNAs.
The Z-scores $z=\frac{x-\mu}{\sigma}$ of purine riboswitches $\alpha$
are highly correlated with the values MFE-EXPB, with $r = 0.9606$.
The right panel of Figure~\ref{fig:Zscores} clearly indicates
that MFE-EXPB for purine riboswitches is about two times larger in absolute
value than than for randomized RNAs.

\section{Discussion}
\label{section:discussion}

Understanding the network topology of macromolecules is important,
both for folding kinetics and for computational approaches to determine the
native structure of protein and RNA.
In this paper, we consider
the network of all secondary structures of a given RNA sequence,
where nodes are connected when the corresponding structures have base
pair distance one. We describe the first algorithm to compute the
expected network degree, where expectation is computed with
respect to either 
the uniform  or Boltzmann probability. Using dynamic programming, our
C-program {\tt RNAexpNumNbors} runs in cubic time with quadratic memory
requirements, although the network size is exponential in input RNA
sequence length.

Our network is related, but different from that of
\cite{Wuchty.nar03}, in which nodes (secondary structures)
$s,t$ are linked by an edge if $t$ can be obtained from $s$ by
adding or removing a base pair, or by by a {\em shift move}.
In contrast to Wuchty's observation, that the
{\em clustering coefficient} $C(\nu)$ for phe-tRNA is inversely proportional
to node degree $k_{\nu}$ of node $\nu$, for our network, $C(\nu)=0$ for
all nodes $\nu$. Indeed, if $s,t$ and $s,u$ are
neighbors, then 
$d_{\mbox{\small BP}}(s,t)=1$ and
$d_{\mbox{\small BP}}(s,u)=1$, so it follows that
$d_{\mbox{\small BP}}(t,u)=2$ -- no two neighbors of $s$ are connected
by an edge.
At present, it is unclear whether
{\tt RNAexpNumNbors} can be extended to allow shift moves, a topic we
hope to explore in future work. 

Using {\tt RNAexpNumNbors}, we analyzed a number of 
noncoding RNAs from
the Rfam database. We have shown that for such RNAs,
the expected degree is generally less than the degree of the 
minimum free energy (MFE)
structure, which in turn is less than the degree of the
minimum free energy structure constrained to be consistent with the
Rfam consensus structure (CMFE).  This observation is consistent with
what one knows from Markov state models formed by clustering structures
from molecular dynamics snapshots \cite{Bowman.pnas10}.
However, the expected degree
of structural RNAs, such as purine riboswitches, paradoxically
appears to be smaller than that of random RNA, yet the difference
between the degree of the MFE structure and the expected degree
is larger than that of random RNA. Expected degree does not
seem to correlate with any structural diversity 
measure of RNA, such as positional entropy, ensemble defect, etc.
Moreover, there is no correlation (-0.054286) between the expected number of 
neighbors and folding time,
nor any corrlation (-0.036524) between the MFE-EXPB and folding time,
as measured by {\tt Kinfold} \cite{flamm}
using a carefully chosen benchmarking set of 
1000 20-mers from \cite{SenterKineticsJCB}.

We close this paper by posing a few questions.
One can argue that the
collection of neighbors of a given structure $s$ constitutes
a conformational {\em breathing} space for thermal movement while retaining
functionality. Hence, the term
$S=k_B \cdot \log N(s)$ is a form of {\em configurational entropy}, 
which is not currently accounted for in RNA secondary structure models.
Can one define the free energy term $c \cdot T \cdot S$,
for absolute temperature $T$, where $c$ is an appropriate weight
with respect to the Turner energy parameters? Would such an additional
energy contribution improve structure prediction? 
Does expected network degree play a role in RNA molecular evolution?
Can the algorithm {\tt RNAexpNumNbors} be extended to allow shift moves,
or to apply to $k$-neighbors, for $k>1$?
Finally, in reference to
Figure~\ref{fig:asymptoticNormalizedExpNumNborsHomopolymer},
can one prove the existence of an asymptotic limit
$\lim_{|\aseq| \rightarrow \infty} \frac{\langle N(\aseq) \rangle}{|\aseq|}$
for homopolymers $\aseq$, using the algebraic
combinatorial techniques of 
\cite{FlSebook,cloteFerreKranakisKrizanc:RNA05,Fusy.jmb12}?

\section{Methods}
\label{section:methods}

In this section, we provide recursions for efficient dynamic programming
algorithms for the expected number $\langle N_s \rangle$ of neighbors of
secondary structure $s$, where $s$ varies over all secondary structures of
a given RNA sequence $\aseq$. For clarity of exposition, we present three
different algorithms, depending on the probability model 
for secondary structures (uniform model versus
Boltzmann with Nussinov energy model versus
Boltzmann with Turner energy model).

\subsection{Model A: uniform probability distribution}

In this subsection, $\langle N_s \rangle$ is formally defined as follows
\begin{eqnarray}
\label{eqn:expectationNx1}
\langle N_s \rangle &=& \frac{\sum_s N_s}{Z}
\end{eqnarray}
where $N_s$ denotes the number of secondary structures, whose base pair
distance with $s$ is $1$, and $Z$ denotes the total number of secondary
structures of given RNA sequence $\aseq$, and the summation is taken over
all secondary structures $s$ of $\aseq$. For any secondary structure $s$,
let $|s|$ denote the number of base pairs in $s$.

Suppose that $\aseq = \aseq_1,\ldots,\aseq_n$. For $1 \leq i \leq j \leq n$,
define $\aseq[i,j] = \aseq_i,\ldots,\aseq_j$, and define $SS(\aseq[i,j])$ to
be the collection of secondary structures of $\aseq[i,j]$. Define
\begin{eqnarray}
\label{eqn:Qij_baseCase1}
Q_{i,j} = \sum_{s \in SS(\aseq[i,j])} N_s.
\end{eqnarray}
Similarly, let $Z_{i,j} = \sum_{s \in SS(\aseq[i,j])} 1$; i.e. $Z_{i,j}$ denotes
the number of secondary structures of $\aseq[i,j]$.
\medskip

\noindent
{\sc Base Case:}
For $j-i \in \{ 0,1,2,3\}$, $Q_{i,j}=0$ and $Z_{i,j}=1$.
\medskip

\noindent
{\sc Inductive Case:} Let $BP(i,j,\aseq)$ be a boolean function, taking
the value $1$ if positions $i,j$ can form a base pair for sequence $\aseq$,
and otherwise taking the value $0$.  Assume that $j-i > 3$.
\medskip

\noindent
{\sc Subcase A:} Consider all secondary structures $s \in \aseq[i,j]$,
for which $j$ is unpaired. For each structure $s$ in this subcase,
the number $N_s$ of neighbors of $s$ is constituted from the number of
structures obtained from $s$ by removal of a single base pair, together
with the number of structures obtained from $s$ by addition of a single
base pair. If the base pair added does not involve terminal position $j$,
then total contribution to $\sum_{s \in SS(\aseq[i,j])} N_s$ is $Q_{i,j-1}$.
It remains to count the contribution due to neighbors $t$ of $s$,
obtained from $s \in SS(\aseq[i,j])$ by adding the base pair $(k,j)$.
This contribution is given by
$\sum_{k=i}^{j-4} BP(k,j,\aseq) \cdot Z_{i,k-1} \cdot Z_{k+1,j-1}$,
where $Z_{i,i-1}$ is defined to be $1$.
Thus the total contribution to $Q_{i,j}$ from this subcase is
\[
Q_{i,j-1} + 
\sum_{k=i}^{j-4} BP(k,j,\aseq) \cdot Z_{i,k-1} \cdot Z_{k+1,j-1}.
\]
\medskip

\noindent
{\sc Subcase B:} Consider all secondary structures $s \in \aseq[i,j]$
that contain the base pair $(k,j)$ for some $k \in \{i,\ldots,j-4\}$.
For secondary structure $s$ in this subcase, the number $N_s$ of neighbors
of $s$ is constituted from the number of structures obtained by removing
base pair $(k,j)$ together with a contribution obtained by adding/removing
a single base pair either to the region $[i,k-1]$ or to the region
$[k+1,j-1]$.
Setting $Q_{i,i-1}$ to be $0$, these contributions are given by 
\[
\sum_{k=i}^{j-4} BP(k,j,\aseq) \cdot \left[
Z_{i,k-1} \cdot Z_{k+1,j-1} +
Q_{i,k-1} \cdot Z_{k+1,j-1} +
Z_{i,k-1} \cdot Q_{k+1,j-1} \right].
\]
In the current subcase, the contribution to $Z_{i,j}$ is
$\sum_{k=i}^{j-4} BP(k,j,\aseq) \cdot Z_{i,k-1} \cdot Z_{k+1,j-1}$.

Finally, taking the contributions from both subcases together, it follows
that
\begin{eqnarray}
\label{eqn:Qij_inductiveCase1}
Q_{i,j} &=& Q_{i,j-1} + 
\sum_{k=i}^{j-4} BP(k,j,\aseq) \cdot \left[ 2 \cdot Z_{i,k-1} \cdot Z_{k+1,j-1} +
Q_{i,k-1} \cdot Z_{k+1,j-1} +
Z_{i,k-1} \cdot Q_{k+1,j-1} \right]\\
\label{eqn:Nij_inductiveCase1}
Z_{i,j} &=& Z_{i,j-1} + 
\sum_{k=i}^{j-4} BP(k,j,\aseq) \cdot Z_{i,k-1} \cdot Z_{k+1,j-1}.
\end{eqnarray}
It follows that the 
expected number $\langle N_s \rangle$ of neighbors $N_s$ of
structures $s$ of $\aseq$ is $\frac{Q_{1,n}}{Z_{1,n}}$.

We should remark that the recursion for $Z_{i,j}$ is well-known and
due originally to Waterman, where in \cite{steinWaterman} the asymptotic
number of secondary structures of a homopolymer is determined. However,
to the best of our knowledge, the recursions and related dynamic programming
algorithm for $\langle N_s \rangle$ are new. We have implemented the 
dynamic programming algorithm corresponding to equations
(\ref{eqn:Qij_baseCase1}) and (\ref{eqn:Qij_inductiveCase1}), as well as
an algorithm proceeding by brute force enumeration as a cross-check of the
first algorithm. Subsequently, we have cross-checked the recursions for
Models B and C by setting energy terms to zero and comparing the results
with our implementation for Model A.

%Finally, we remark that it is straightforward to use algebraic
%combinatorial methods, as in \cite{Clote.jbcb09,Fusy.jmb12} to determine
%the asymptotic value of $\langle N_s \rangle$ for the homopolymer model
%of RNA. We do not go into details, since this has no direct bearing on 
%the kinetics of RNA folding.
%In the homopolymer model, the recursions are:
%Q(n) = 
%Q(n-1) + sum_{k=0}^{n-6} 2*N(k)*N(n-2-k) + Q(k)*N(n-2-k) + N(*k)*Q(n-2-k)
%where Q(-1)=Q(0)=Q(1)=Q(2)=Q(3)=Q(4)=0
%asymptotics ASYMPTOTICS Asymptotics

Now we give the recursions for a dynamic programming algorithm
to compute
$\langle N \rangle = \sum_{s} P(s) \cdot N(s)$, where the sum is taken over
all secondary structures $s$ of the input RNA sequence ${\bf a} = 
a_1,\ldots,a_n$,
$N(s)$ is the number of structures of ${\bf a}$ that differ by one base pair
from $s$, and
$P(s) = \sum_{\exp(-E(s)/RT)}$ is the Boltzmann probability of structure $s$,
where $E(s)$ is alternately the Nussinov base pairing energy model or the
Turner base stacking energy model. We provide derivations for the recursions
in the Appendix.

\subsection*{Model B: Base pairing energy}

Here we consider the Nussinov energy model \cite{nussinovJacobson},
where each base pair of a secondary structure contributes an energy of $-1$.
It follows that for secondary structure $s$ of $\aseq=\aseq_1,\ldots,\aseq_n$,
$E(s) = -1 \cdot |s|$, where $|s|$ denotes the 
number of base pairs in $s$.  For this model, the expected number of
neighbors $\langle N_s \rangle$ is defined by
\begin{eqnarray}
\label{eqn:expectationNx2}
\langle N_s \rangle &=&
\frac{\sum_{s \in SS(\aseq)} N_s \cdot \exp(-E(s)/RT)}{Z} =
\sum_{s \in SS(\aseq)} N_s \cdot P(s)\\
\end{eqnarray}
where $P(s)=\frac{\exp(-E(s)/RT)}{Z}$ 
denotes the Boltzmann probability of structure $s$.
(In the previous section, the uniform probability of $s$ was $1/Z$,
where $Z$ denoted the number of structures.)
In contrast to the previous subsection,
here we define $Q,Z$ as follows
\begin{eqnarray}
\label{eqn:Qij_defNJ}
Q_{i,j} &=& \sum_{s \in SS(\aseq[i,j])} N_s \cdot \exp(-E(s)/RT)\\
Z_{i,j} &=& \sum_{s \in SS(\aseq[i,j])} \exp(-E(s)/RT). \nonumber
\end{eqnarray}
\medskip

In \cite{nussinovJacobson}, the energy function $E_{i,j}$ is defined to be
$-1$ if positions $i,j$ can form a base pair, and otherwise $E_{i,j}=0$.
A slightly better refinement is the following energy function that 
one could adopt:
\begin{eqnarray*}
E_{i,j} = \left\{ \begin{array}{ll}
-3 &\mbox{if $a_i,a_j = GC$ or $a_i,a_j = CG$}\\
-2 &\mbox{if $a_i,a_j = AU$ or $a_i,a_j = UA$}\\
-1 &\mbox{if $a_i,a_j = GU$ or $a_i,a_j = UG$}\\
0 &\mbox{otherwise}.
\end{array} \right.
\end{eqnarray*}

\noindent
{\sc Base Case:}
For $1 \leq i \leq n$ and $j=i-1$, define $Z_{i,j}=1$ and $Q_{i,j}=0$.
For $j \in \{i,\ldots,i+3\}$, define $Z_{i,j}=1$ and
$ZB_{i,j}=ZM_{i,j}=ZM1_{i,j}=0$ and $QB_{i,j}=QM_{i,j}=QM1_{i,j}=0$. 
\medskip

\noindent
{\sc Inductive Case:} For $i+3<j$,
\begin{eqnarray*}
%New incorrect recursion below:
%Q_{i,j} &=&Q_{i,j-1}+ \sum_{k=i}^{j-4} bp(k,j) \cdot 
%        \exp\left(\frac{-E_{k,j}}{RT} \right) \cdot
%       \left\{ Z_{i,k-1} Z_{k+1,j-1} \right.
%+ \\
%&& \left. Q_{i,k-1} Z_{k+1,j-1} + Z_{i,k-1} Q_{k+1,j-1}  \right\} \\
%Z_{i,j} &=&Z_{i,j-1} + \sum_{k=i}^{j-4} bp(k,j) \cdot
%\exp\left(\frac{-E_{k,j}}{RT} \right) \cdot Z_{i,k-1} \cdot Z_{k+1,j-1}.
%
%Original recursion which agrees with appendix:
Q_{i,j} &=&Q_{i,j-1}+ \sum_{k=i}^{j-4} bp(k,j) \cdot \left[
Z_{i,k-1} Z_{k+1,j-1}
\left( 1 + \exp\left(\frac{-E_{k,j}}{RT} \right) \right) \right] 
+ \\
&& \sum_{k=i}^{j-4} bp(k,j) \cdot 
\exp\left(\frac{-E_{k,j}}{RT} \right) \cdot \left[
Q_{i,k-1} Z_{k+1,j-1} + Z_{i,k-1} Q_{k+1,j-1}  \right] \\
Z_{i,j} &=&Z_{i,j-1} + \sum_{k=i}^{j-4} bp(k,j) \cdot
\exp\left(\frac{-E_{k,j}}{RT} \right) \cdot Z_{i,k-1} \cdot Z_{k+1,j-1}.
\end{eqnarray*}

\subsection*{Model C: Turner energy}
Define the following helper functions:
\begin{eqnarray*}
arc1(i,j) &=& |\{ (x,y): bp(x,y)=1, i \leq x<y \leq j, x+3<y \}| \\
arc2(i,j,\ell,r) &=& |\{ (x,y): bp(x,y)=1, i < x < \ell,
r < y < j  \}| \\
arc3(i,j,\ell,r) &=& arc1(i+1,\ell-1)+arc1(r+1,j-1) + arc2(i,j,\ell,r).
\end{eqnarray*}
Note the occurrence of inequality $\leq$ in $arc1$, in contrast to the
occurrence of strict inequality $<$ in $arc2$. Clearly, $arc1(i,j)$ is
the number of potential base pairs in the input
RNA sequence $a_1,\ldots,a_n$ that are found in the interval $[i,j]$.
In contrast, $arc2(i,j,\ell,r)$ is the number of potential base pairs
$(x,y)$, where $x$ occurs in the left bulge and $y$ occurs in the right
bulge of a reference structure; i.e. the number of potential base pairs
that `bridge' an internal loop. Finally, $arc3(i,j,\ell,r)$ is the number
of potential base pairs occurring in the left bulge, right bulge or
`bridging' the internal loop. Of course, it is possible that $\ell = i+1$
[resp. $r = j-1$], in which case there is no left bulge [resp. right bulge]
and hence no internal loop.
\medskip

\noindent
{\sc Base Case:}
For $1 \leq i \leq n$ and $j=i-1$, define $Z_{i,j}=1$ and $Q_{i,j}=0$.
For $j \in \{i,\ldots,i+3\}$, define $Z_{i,j}=1$ and
$ZB_{i,j}=ZM_{i,j}=ZM1_{i,j}=0$ and $QB_{i,j}=QM_{i,j}=QM1_{i,j}=0$. 
\medskip

\noindent
{\sc Inductive Case:} For $i+3<j$, 
\begin{eqnarray*}
Q_{i,j} &=&Q_{i,j-1}+ \sum_{k=i}^{j-4} bp(k,j) \cdot \left[
Z_{i,k-1}Z_{k+1,j-1} + Q_{i,k-1}ZB_{k,j} + Z_{i,k-1}QB_{k,j} \right]\\
QB_{i,j} &=& A_{i,j}+B_{i,j} + C_{i,j}\\
A_{i,j} &=& \exp\left( -\frac{H(i,j)}{RT} \right) \cdot \left[ 1+ 
arc1(i+1,j-1) \right] \\
B_{i,j} &=& \sum_{\ell=i+1}^{\min(i+31,j-5)} \sum_{r=j-1}^{\max(j-31,i+5)}
\exp\left( -\frac{I(i,j,\ell,r)}{RT} \right) \cdot \left[ 
ZB_{\ell,r}\left(1 + arc3(i,j,\ell,r)\right) + QB_{\ell,r} \right]\\
C_{i,j} &=& \sum_{r=i+5}^{j-5} \exp\left( -\frac{a+b}{RT} \right) \cdot 
\left[ QM_{i+1,r-1}ZM1_{r,j-1} + ZM_{i+1,r-1}QM1_{r,j-1} \right] \\
QM1_{i,j} &=& \sum_{k=i+4}^{j} \exp\left( -\frac{c(j-k)}{RT} \right) \cdot 
\left[ QB_{i,k}+ZB_{i,k} \cdot arc1(k+1,j) \right] \\
QM_{i,j} &=& \sum_{r=i}^{j-5} \exp\left( -\frac{b+c(r-i)}{RT} \right) \cdot 
QM1_{r,j} + ZM1_{r,j} \cdot arc1(i,r-1) + \\
&& \sum_{r=i}^{j-5} 
%\exp\left( -\frac{b+c(r-i)}{RT} \right) \cdot 
\left[ QM_{i,r-1} ZM1_{r,j}  + ZM_{i,r-1} QM1_{r,j} \right]
\end{eqnarray*}

Finally, to accelerate the computation of the functions $arc1,arc2$,
the $4\times n \times n$ array $ARC$ should be precomputed, where if 
${\bf a} = a_1,\ldots,a_n$ denotes the input RNA sequence, then
\begin{eqnarray*}
ARC[\alpha,i,j] = \left\{ \begin{array}{ll}
|{ x \in [i,j] : a_x = U }|
 &\mbox{if $\alpha=0$}\\
|{ x \in [i,j] : a_x = G }|
 &\mbox{if $\alpha=1$}\\
|{ x \in [i,j] : a_x \in \{C,U\} }|
 &\mbox{if $\alpha=2$}\\
|{ x \in [i,j] : a_x \in \{A,G\} }|
 &\mbox{if $\alpha=3$}. \end{array} \right.
\end{eqnarray*}
If $index(\alpha)=0,1,2,3$ respectively for values $\alpha=A,C,G,U$,
then 
$arc1(i,j) = \sum_{k=i}^{j-4} ARC[index(a_{k}),k+4,j]$ and 
$arc2(i,j,\ell,r) = \sum_{k=i+1}^{\ell-1} ARC[index(a_{k}),r+1,j-1]$.

Note that in the implementation of $B_{i,j}$, the first sum
$\sum_{\ell=i+1}^{\min(i+31,j-5)}$ is implemented by the FOR loop
\begin{quote}
for $\ell=i+1$ to $\min(i+31,j-5)$
\end{quote}
while the second sum $\sum_{r=j-1}^{\max(j-31,i+5)}$
is implemented by the reverse FOR loop
\begin{quote}
for $r=j-1$ down to $\max(j-31,i+5)$
\end{quote}
and although not written explicitly in the expression for $B_{i,j}$,
there is a check that $(\ell-i-1) + (j - r -1) \leq 30$. This follows
the convention in Vienna RNA Package that internal loops have size bounded
by 30.

It is worth noting that if all energy terms are set to zero, then
$Q_{i,j}$ in this section is {\em not} necessarily equal to $N_{i,j}$, in the
treatment of the uniform probability case. This is because we have
ignored structural neighbors formed by addition of a base pair $(x,y)$
in a multiloop structure $s$ closed by base pair $(i,j)$, where
$i< x< \ell < r  < y < j$, while $(\ell,r) \in s$; i.e. the base pair
$(x,y)$ spans one or more components of a multiloop and connects  the
previously unpaired positions $x,y$ in the multiloop $s$. We are obliged
to ignore such potential structural neighbors because of technical treatment
of multiloops in the McCaskill partition function \cite{mcCaskill}.
Nevertheless, when using the Turner energy parameters,
in practice there should be only a small discrepancy with the true value of
$Q_{i,j}$ as computed by brute force. This is because we expect both the
number of unpaired bases and the number of components in a multiloop to 
be small, so there will be few occasions where this special case might
arise (otherwise, this is energetically unfavorable, and hence the
Boltzmann probability would be small).

\section{Acknowledgements}

The work reported in this paper was done during a visit with
Niles Pierce at the California Institute of Technology,
Knut Reinert at the Free University of Berlin and
Martin Vingron at the Max Planck Institute for Molecular Genetics.
Warm thanks are due to all three persons. We would like to thank
the reviewers for their very helpful comments.
This research was funded by a Guggenheim Fellowship,
National Science Foundation grant DBI-1262439, and funding from the
Deutscher Akademischer Austauschdienst.  Any opinions, findings,
and conclusions or recommendations expressed in this material are
those of the authors and do not necessarily reflect the views of the
National Science Foundation.

\hfill\break\newpage \clearpage

\section*{Appendix}
\label{section:appendix}

\section*{1. RNA structural measures}

In the main text, we determined the correlation between the expected
network degree and the following RNA structural measures:
expected number of base pairs, expected base pair distance,
ensemble defect, positional entropy. In our correlations in the main
text, each of these measures was length-normalized  by dividing the
value by sequence length.
These structural measures can be defined from the base pairing probabilities,
computed by McCaskill's algorithm \cite{mcCaskill} and implemented in
{\tt RNAfold -p} \cite{hofacker:ViennaWebServer}.  Let
\begin{eqnarray}
\label{def:pij}
p_{i,j} = \sum_{\{ s: (i,j) \in s\}} P(s) =
\frac{\sum_{\{ s: (i,j) \in s\}} \exp(-E(s)/RT) }{Z} 
\end{eqnarray}
where $P(s)$ is the Boltzmann probability of structure $s$ of a given
RNA sequence  $\aseq=\aseq_1,\ldots,\aseq_n$,
$E(s)$ is the Turner 1999 energy of secondary 
structure $s$ \cite{turner,xia:RNA},
$R \approx 0.001987$ kcal mol$^{-1}$ K$^{-1}$ is the universal gas 
constant, $T = 310.15$ is
absolute temperature, and the {\em partition function}
$Z = \sum_{s} \exp(-E(s)/RT)$, where the sum is taken over all
secondary structures $s$ of $\aseq$. Symmetrize the base pair probabilities,
by defining $p_{j,i}=p_{i,j}$ for $1 \leq i,j \leq n$, $i \ne j$, 
and define $p_{i,i} = 1- \sum_{i \ne j} p_{i,j}$ to be the probability
that position $i$ is unpaired. Let $s_0$ denote the minimum free energy
structure of input RNA sequence $\aseq$.

\begin{enumerate}
\item
Expected number of base pairs (ExpNumBP) is defined by
$\sum_{1 \leq i<j \leq n} p_{i,j}$.

\item
Expected base pair distance (ExpBPDist) to the MFE structure
$s_0$ of input RNA sequence $\aseq$ is defined by
$\sum_{1 \leq i<j \leq n}
I[ (i,j) \not\in s_0 ] \cdot p_{i,j} +
I[ (i,j) \in s_0 ] \cdot (1-p_{i,j}),
$
where $I$ denotes the indicator function. 

\item
Ensemble defect (EnsDef)
is the expected number of nucleotides whose base pairing status differs
from the MFE structure $s_0$,
defined by
$n - \sum_{i \ne j} p_{i,j} \cdot I[ (i,j) \in s_0 ]
- \sum_{1 \leq i \leq n} p_{i,i}\cdot I[ \mbox{$i$ unpaired in $s_0$}]$,
where $I$ is the indicator function.  

\item
Total positional entropy (H) is defined by
$\langle H(\aseq) \rangle = \sum_{i=1}^n 
\left\{  - \left( p_{i,i} \cdot \ln p_{i,i} + (1- p_{i,i}) \cdot 
\ln (1-p_{i,i}) 
\right) \right\}$, where $0 \cdot \ln 0$ is defined to be $0$.
\end{enumerate}

\section*{2. Expected number of neighbors for Boltzmann distribution}

In this section, we provide full details on the derivation of the recursions
from Section~\ref{section:methods}. 
%This section is technical and would
%require a lengthy explanation of many terms, such as multiloop, multiloop
%with one component, etc. to be self-contained. In an already long paper,
%the additional length cannot be justified, and we must assume that the
%reader is familiar with the definitions and constructions in McCaskill's
%paper \cite{mcCaskill}.
%
By setting all energy terms to zero
in the recursions for Models A, B and C, we should obtain the same value
as in the uniform probability case. In testing {\tt RNAexpNumNbors}, this
is indeed the case, except for a very slight undercount in multiloops
in Model C. For the reasons explained at the end of the Appendix, this
will make little difference when using the Turner energy parameters, since
multiloops are energetically unfavorable.

Throughout this section, ${\bf a} = a_1,\cdots,a_n$ denotes an arbitrary
but fixed RNA sequence.  Below, we justify the recursions
given for $\langle Q(s) \rangle = \sum_{s} BF(s) \cdot N(s)$, where the sum is 
taken over all secondary structures $s$ of RNA sequence ${\bf a}$,
$N(s)$ is the number of structures of ${\bf a}$ that differ by one base pair
from $s$, and the Boltzmann factor $BF(s)$ of $s$ is defined by
$\exp(-E(s)/RT)$, where $E(s)$ is the free energy of $s$.
Recursions are also given for the 
partition function $Z(s) = \sum_s \exp(-E(s)/RT)$, where the sum is taken
over all secondary structures of ${\bf a}$. It follows that
the expected number of structural neighbors 
\[
\langle N \rangle = \sum_{s} BF(s) \cdot N(s) = \frac{Q(s)}{Z(s)}
\]
For $1\leq i\leq j \leq n$, the collection of
all secondary structures of $\aseq[i,j] = a_i,\ldots,a_j$ is denoted $ss[i,j]$. In contrast,
if $s$ is a secondary structure of $a_1,\ldots,a_n$, then $s[i,j]$ is the
{\em restriction} of $s$ to the interval $[i,j]$, defined by
$s[i,j] = \{ (x,y) : i \leq x \leq y \leq j, (x,y) \in s$.

We give separate algorithms for the expected number $\langle N \rangle$
of structural neighbors, depending on whether the free energy
$E(s)$ is computed with respect to the Nussinov base pairing energy model or the
Turner base stacking energy model. However, the initial portion of the
derivation is common to both energy models. Define 
\begin{eqnarray*}
QB_{i,j}&=& \sum_{\substack{s \in ss[i,j]\\ \text{$(i,j) \in s$}}} BF(s)N(s) \\
ZB_{i,j}&=& \sum_{\substack{s \in ss[i,j]\\ \text{$(i,j) \in s$}}} BF(s). 
\end{eqnarray*}

\subsection*{1.1 Initial derivation shared by energy Model B and C}

For notational convenience, we define $Q_{i,i-1}=0$ and $Z_{i,i-1}=1$.
If $i \leq j < i+4$,  then for any secondary structure $s$, 
there are no structural neighbors of $s[i,j]$ and so $Q_{i,j} = 0$.
As well, the only secondary structure on $[i,j]$ is the empty structure,
so $Z_{i,j}=1$. 

Now assume that $i+4 \leq j$. Since
\[
Q_{i,j}= \sum_{\substack{s \in ss[i,j]\\ \text{$j$ unpaired in $s$}}} 
BF(s) N(s) + \sum_{k=i}^{j-4} 
\sum_{\substack{s \in ss[i,j]\\ \text{$(k,j) \in s$}}} 
BF(s) N(s).
\]
we treat each sum in a separate case. Let $bp(k,j)$ be a boolean valued
function with the value $1$ if $k$ can base-pair with $j$; i.e. 
$a_ka_j \in \{ AU,UA,CG,GC,GU,UG \}$. For secondary structure $s\in ss[i,j]$,
let $bp(k,j,s)$ be a boolean function with value $1$ if it is possible to
add the base pair $(k,j)$ to $s$ and obtain a valid secondary structure; i.e.
without creating a base triple or pseudoknot.
\medskip

\noindent
{\sc Case 1:} $j$ is unpaired in  $[i,j]$. For $s \in ss[i,j]$ in which
$j$ is unpaired, $s=s[i,j-1]$ and $BF(s)=BF(s[i,j-1])$. The contribution to
$Q_{i,j}$ in this case is given by
\begin{eqnarray*}
A_{i,j} &=& \sum_{\substack{s \in ss[i,j]\\ \text{$j$ unpaired in $s$}}} 
BF(s) N(s) \\
&=& \sum_{\substack{s \in ss[i,j-1]\\ \text{~}}} BF(s) \left[ N(s) 
+ \sum_{\substack{\text{$i\leq k \leq j-4$, $k$ unpaired in $s$}\\ \text{$bp(k,j,s)=1$}}}  
1 \right] \\
&=&Q_{i,j-1} + \sum_{k=i}^{j-4} bp(k,j) 
%\sum_{s_1 \in ss[i,k-1]} \sum_{\substack{\text{$s_2 \in ss[k,j]$}\\
%\text{$(k,j) \in s_2$}}}
\sum_{s_1 \in ss[i,k-1]} \sum_{s_2 \in ss[k+1,j-1]}\\
&& BF(s_1) \cdot BF(s_2) \\
&=&Q_{i,j-1} + \sum_{k=i}^{j-4} bp(k,j) \cdot Z_{i,k-1} \cdot Z_{k+1,j-1}\\
\end{eqnarray*}
The term $1$ on the right side of line 2 arises from neighbors of $s$
obtained by adding the base pair $(k,j)$ to $s$. In line 3, note that
if $s \in ss[i,j]$ is a structure in which both $k,j$ are unpaired and
$bp(k,j,s)=1$, then $BF(s) = BF(s_1) \cdot BF(s_2)$.
\medskip

\noindent
{\sc Case 2:} $j$ is paired in  $[i,j]$.  The contribution to
$Q_{i,j}$ in this case is given by
\begin{eqnarray*}
B_{i,j} &=& \sum_{k=i}^{j-4} \sum_{\substack{s \in ss[i,j]\\ \text{$(k,j) \in s$}}} 
BF(s) N(s) 
= \sum_{k=i}^{j-4} \sum_{\substack{s \in ss[i,j]\\ \text{$(k,j) \in s$}}} 
BF(s) \left[ N(s[i,k-1]) + N(s[k,j]) \right]\\
&=& \sum_{k=i}^{j-4} bp(k,j) \cdot \left\{ 
\sum_{\substack{s_1 \in ss[i,k-1]\\ \text{~}}} 
\sum_{\substack{s_2 \in ss[k,j]\\ \text{$(k,j) \in s_2$}}} 
BF(s_1)\cdot BF(s_2)  \left[ N(s_1) + N(s_2) \right] \right\}\\ 
&=& \sum_{k=i}^{j-4}  bp(k,j) \cdot \left\{
\sum_{\substack{s_1 \in ss[i,k-1]\\ \text{~}}} BF(s_1) N(s_1) 
\sum_{\substack{s_2 \in ss[k,j]\\ \text{$(k,j)\in s_2$}}} BF(s_2) + \right. \\
&&\left. \sum_{\substack{s_1 \in ss[i,k-1]\\ \text{~}}} BF(s_1) 
\sum_{\substack{s_2 \in ss[k,j]\\ \text{$(k,j)\in s_2$}}} BF(s_2) N(s_2) 
\right\}  \\
&=& \sum_{k=i}^{j-4}  bp(k,j) \cdot \left\{
Q_{i,k-1} \cdot ZB_{k,j} +
Z_{i,k-1} \cdot QB_{k,j} \right\}. 
\end{eqnarray*}
Putting together the contributions from both cases, we have
\begin{eqnarray}
\label{eqn:nussTurnerExpNhbors}
Q_{i,j} &=& Q_{i,j-1} + \sum_{k=i}^{j-4} bp(k,j) \left[
Z_{i,k-1} Z_{k+1,j-1} +
Q_{i,k-1} ZB_{k,j} +
Z_{i,k-1} QB_{k,j}  \right].
\end{eqnarray}

\subsection*{1.2 Model B: Base pairing energy}
In the Nussinov energy model, where the energy of base pair $(k,j)$ is
denoted $E(k,j)$, we clearly have the following.
\begin{eqnarray*}
QB_{i,j} &=& 
\sum_{\substack{s \in ss[i,j]\\ \text{$(i,j)\in s$}}} BF(s) N(s)\\ 
&=&
\sum_{\substack{s \in ss[i,j]\\ \text{$(i,j)\in s$}}} BF(s) \left[
1 + N(s[i+1,j-1]) \right] \\ 
&=&ZB_{i,j} + \exp(-E(i,j)/RT) \cdot Q_{i+1,j-1}\\
ZB_{i,j} &=& \exp(-E(i,j)/RT) \cdot Z_{i+1,j-1}.
\end{eqnarray*}
The term $1$ arises from those neighbors of $s$, obtained by removal of
the base pair $(i,j)$ while the term $N(s[i+1,j-1])$ arises from neighbors
of $s$ obtained by removal/addition of a base pair within $[i+1,j-1]$.
We now substitute the expressions for $QB_{k,j}$ and $ZB_{k,j}$ into
equation (\ref{eqn:nussTurnerExpNhbors}).
\begin{eqnarray*}
\label{eqn:nussExpNhbors}
Q_{i,j} &=& Q_{i,j-1} + \sum_{k=i}^{j-4} bp(k,j) \left[
Z_{i,k-1} Z_{k+1,j-1} +
Q_{i,k-1} ZB_{k,j} +
Z_{i,k-1} QB_{k,j}  \right] \noindent \\
&=& Q_{i,j-1} + \sum_{k=i}^{j-4} bp(k,j) \left\{
Z_{i,k-1} Z_{k+1,j-1} +
Q_{i,k-1} Z_{k+1,j-1} \cdot \exp\left( - \frac{E(k,j)}{RT} \right) \right. +\\
&& Z_{i,k-1} \left. \left[ 
Z_{k+1,j-1} \cdot \exp\left( - \frac{E(k,j)}{RT} \right) + 
Q_{k+1,j-1} \cdot \exp\left( - \frac{E(k,j)}{RT} \right)  \right] \right\} \\
&=& Q_{i,j-1} + \sum_{k=i}^{j-4} bp(k,j) \left\{
Z_{i,k-1} Z_{k+1,j-1} \left( 1 + \exp\left( - \frac{E(k,j)}{RT} \right) 
\right) \right. +
\\
&&
\left. \exp\left( - \frac{E(k,j)}{RT} \right) \cdot \left[
Q_{i,k-1}Z_{k+1,j-1} + 
Z_{i,k-1}Q_{k+1,j-1}  \right] \right\}
\end{eqnarray*}
Note that if we set all base pair energies 
$E(k,j)$ to $0$, then we obtain the same expression as derived for
the uniform probability distribution.

\begin{figure*}[!ht]
\begin{center}
\includegraphics[width=.8\linewidth]{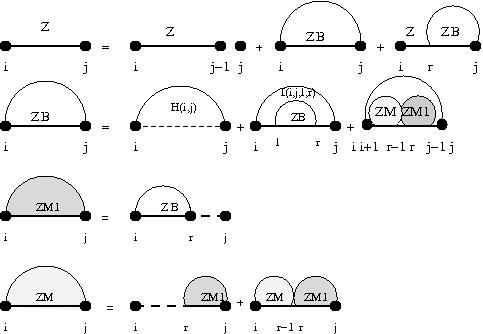}
\caption{Feynman diagram which illustrates the recursions for
McCaskill's algorithm \cite{mcCaskill}.
}
\label{fig:feynmanDiagram}
\end{center}
\end{figure*}

\subsection*{1.3 Model C: Turner nearest neighbor energy model}

In the nearest neighbor energy model 
\cite{zukerMathewsTurner:Guide,Turner.nar10}, free energies are
defined not for base pairs, but rather for {\em loops} in the loop
decomposition of a secondary structure. In particular, there are
stabilizing, negative free energies for stacked base pairs and
destabilizing, positive free energies for hairpins, bulges, internal loops,
and multiloops. 

In this section, free energy parameters for base stacking and loops
are from the Turner 2004 energy model \cite{Turner.nar10}.
As in the previous subsection, $Q,Z$ are defined, but now with respect 
to the Turner model, rather than the Nussinov model.
\begin{eqnarray}
\label{eqn:Qij_defTurner}
Q_{i,j} &=& \sum_{s \in ss[i,j]} N_s \cdot \exp(-E(s)/RT)\\
Z_{i,j} &=& \sum_{s \in ss[i,j]} \exp(-E(s)/RT). \nonumber
\end{eqnarray}
It follows that
$Z=Z_{1,n}$ is the partition function for secondary structures
(the Boltzmann weighted counting of all structures of $\aseq$)
and
\begin{eqnarray}
\label{eqn:NxBoltzmann}
\langle N_s \rangle =
\frac{Q_{1,n}}{Z_{1,n}} = \sum_{s \in ss[1,n]} N_s \cdot P(s) =
\sum_{s \in ss[1,n]} N_s \cdot \frac{\exp(-E(s)/RT)}{Z}.
\end{eqnarray}
To complete the derivation of recursions for $\langle N_s \rangle$,
we must define $QB_{i,j}$ and $ZB_{i,j}$ in
equation (\ref{eqn:nussTurnerExpNhbors}) for the Turner model.

To provide a self-contained treatment, 
we recall McCaskill's algorithm \cite{mcCaskill}, which efficiently computes
the partition function.
For RNA nucleotide sequence $\aseq = \aseq_1,\ldots,\aseq_n$, let
$H(i,j)$ denote the free energy of a hairpin closed by
base pair $(i,j)$, while
$IL(i,j,i',j')$ denotes the free energy of an {\em internal loop}
enclosed by the base pairs $(i,j)$ and $(i',j')$, where $i<i'<j'<j$.
Internal loops comprise the cases of
stacked base pairs, left/right bulges and proper internal loops.
The free energy for a multiloop containing 
$N_b$ base pairs and $N_u$ unpaired bases 
is given by the affine approximation $a+bN_b+cN_u$.

\begin{definition}[Partition function $Z$ and related function $Q$]
 \hfill\break
\label{def:partitionFunctionDefMcCaskill}
\begin{itemize}
\item
$Z_{i,j} = \sum_{s} \exp(-E(s)/RT)$ where the sum is taken over
all structures $s \in ss[i,j]$.
\item
$ZB_{i,j} = \sum_{s}  \exp(-E(s)/RT)$ where the sum is taken over
all structures $s \in ss[i,j]$ which contain the base pair $(i,j)$.
\item
$ZM_{i,j} = \sum_{s} \exp(-E(s)/RT)$ where the sum is taken over
all structures $s \in ss[i,j]$ which are contained
within an enclosing multiloop having {\em at least} one component.
\item
$ZM1_{i,j} = \sum_{s}  \exp(-E(s)/RT)$ where the sum is taken over
all structures $s \in ss[i,j]$ which are contained
within an enclosing multiloop having {\em exactly} one component.
Moreover, it is {\em required} that $(i,r)$ is a base pair of $x$,
for some $i<r\leq j$.
\item
$Q_{i,j} = \sum_{s} N_s \cdot \exp(-E(s)/RT)$ where the sum is taken over
all structures $s \in ss[i,j]$.
\item
$QB_{i,j} = \sum_{s} N_s \cdot \exp(-E(s)/RT)$ where the sum is taken over
all structures $s \in ss[i,j]$ which contain the base pair $(i,j)$.
\item
$QM_{i,j} = \sum_{s} N_s \cdot \exp(-E(s)/RT)$ where the sum is taken over
all structures $s \in ss[i,j]$ which are contained
within an enclosing multiloop having {\em at least} one component.
\item
$QM1_{i,j} = \sum_{s}  N_s \cdot \exp(-E(s)/RT)$ where the sum is taken over
all structures $s \in ss[i,j]$ which are contained
within an enclosing multiloop having {\em exactly} one component.
Moreover, it is {\em required} that $(i,r)$ is a base pair of $s$,
for some $i<r\leq j$.
\end{itemize}
\end{definition}

For $j-i \in \{0,1,2,3\}$, $Z(i,j)=1$, since the empty structure is the only
possible secondary structure. for $j-i>3$, we have
\begin{eqnarray*}
Z_{i,j} &= &Z_{i,j-1} + ZB_{i,j} + 
 \sum_{r=i+1}^{j-4} Z_{i,r-1} \cdot ZB_{r,j} \\
ZB_{i,j} &= &\exp(-HP(i,j)/RT) + 
\displaystyle\sum_{i \leq \ell \leq r \leq j}
\exp(-IL(i,j\ell,r)/RT)\cdot ZB_{{\ell},r} +\\
&& \exp(-(a+b)/RT) \cdot \left( \sum_{r=i+1}^{j-\theta-2} ZM_{i+1,r-1}
\cdot ZM1_{r,j-1} \right)\\
ZM1_{i,j} &= &
\displaystyle\sum_{r=i+\theta+1}^j ZB_{i,r} \cdot
\exp(-c(j-r)/RT) \\
ZM_{i,j} &= &
\displaystyle\sum_{r=i}^{j-\theta-1}  ZM1_{(r,j} \cdot
\exp(-(b+c(r-i))/RT)  + \\
&&\displaystyle\sum_{r=i+\theta+2}^{j-\theta-1}  ZM_{i,r-1} \cdot
ZM1_{r,j} \cdot \exp(-b/RT) . 
\end{eqnarray*}
See Figure \ref{fig:feynmanDiagram} for a pictorial representation
of the recursions of McCaskill's algorithm \cite{mcCaskill},
\medskip

\noindent
{\sc Base Case:}
For $j-i \in \{ -1,0,1,2,3\}$, 
$Q_{i,j}= QB_{i,j}=0$, 
$Z_{i,j}=1$, $ZB_{i,j}= ZM_{i,j}= ZM1_{i,j}=0$. 
\medskip

\noindent
{\sc Inductive Case:} Assume that $j-i > 3$.  Define
\begin{eqnarray*}
arc1(i,j) &=& |\{ (x,y): bp(x,y)=1, i \leq x<y \leq j, x+3<y \}| \\
arc2(i,j,\ell,r) &=& |\{ (x,y): bp(x,y)=1, i < x < \ell <
r < y < j  \}| \\
arc3(i,j,\ell,r) &=& arc1(i+1,\ell-1)+arc1(r+1,j-1) + arc2(i,j,\ell,r).
\end{eqnarray*}
%Note the occurrence of inequality $\leq$ in $arc1$, in contrast to the
%occurrence of strict inequality $<$ in $arc2$. Clearly, $arc1(i,j)$ is
%the number of potential base pairs in the input
%RNA sequence $a_1,\ldots,a_n$ that are found in the interval $[i,j]$.
%In contrast, $arc2(i,j,\ell,r)$ is the number of potential base pairs
%$(x,y)$, where $x$ occurs in the left bulge and $y$ occurs in the right
%bulge of a reference structure; i.e. the number of potential base pairs
%that `bridge' an internal loop. Finally, $arc3(i,j,\ell,r)$ is the number
%of potential base pairs occurring in the left bulge, right bulge or
%`bridging' the internal loop. Of course, it is possible that $\ell = i+1$
%[resp. $r = j-1$], in which case there is no left bulge [resp. right bulge]
%and hence no internal loop.
\medskip

\noindent
{\sc Case A:} $(i,j)$ closes a hairpin.
\medskip

In this case, the contribution to $QB_{i,j}$ is given by 
\begin{eqnarray*}
A_{i,j} &=& \exp \left( - \frac{H(i,j)}{RT} \right)\cdot
\left[ 1 + arc1(i+1,j-1) \right].
\end{eqnarray*}
The term $1$ arises from the neighbor of $s = \{ (i,j) \}$ 
by removing base pair $(i,j)$. The term $arc1(i,j)$ arises from
neighbors of $s$ obtained by adding a base pair in the region $[i+1,j-1]$.
\medskip

\noindent
{\sc Case B:} $(i,j)$ closes a stacked base pair, bulge or internal loop,
whose other closing base pair is $(\ell,r)$, where $i<\ell<r<j$.
\medskip

In this case, the contribution to $QB_{i,j}$ is given by the following
\begin{eqnarray*}
B_{i,j} &=& \sum_{\ell=i+1}^{\min(i+31,j-5)}
\sum_{r=j-1}^{\max(j-31,i+5)}
\exp \left( - \frac{IL(i,j,\ell,r)}{RT} \right)\cdot
\sum_{\substack{s \in ss[\ell,r]\\ \text{$(\ell,r)\in s$}}} BF(s) \left[
1 + arc3(i,j,\ell,r)+N(s) \right] \\
&=& \sum_{\ell=i+1}^{\min(i+31,j-5)} \sum_{r=j-1}^{\max(j-31,i+5)}
\exp \left( - \frac{IL(i,j,\ell,r)}{RT} \right)\cdot
ZB_{\ell,r} \cdot \left[1 + arc3(i,j,\ell,r) \right] + \\
&& \sum_{\ell=i+1}^{\min(i+31,j-5)} \sum_{r=j-1}^{\max(j-31,i+5)}
\exp \left( - \frac{IL(i,j,\ell,r)}{RT} \right)\cdot
QB_{\ell,r}\\
&=& \sum_{\ell=i+1}^{\min(i+31,j-5)} \sum_{r=j-1}^{\max(j-31,i+5)}
\exp \left( - \frac{IL(i,j,\ell,r)}{RT} \right)\cdot
\left[ZB_{\ell,r} \cdot \left( 1 + arc3(i,j,\ell,r) \right) +
QB_{\ell,r} \right].
\end{eqnarray*}
In the summation notation $\displaystyle\sum_{i=a}^b$, 
if upper bound $b$ is smaller than lower bound $a$, then we intend
a loop of the form: FOR $i=b$ downto $a$.
\medskip

\noindent
{\sc Case C:} $(i,j)$ closes a multiloop.
\medskip

In this case, the contribution to $QB_{i,j}$ is given by the following
\begin{eqnarray*}
C_{i,j} &=& 
\sum_{\substack{s \in ss[i,j], (i,j)\in s\\ \text{$(i,j)$ closes a multiloop}}} BF(s)N(s)\\ 
&=& 
\sum_{r=i+5}^{j-5}
\exp \left( - \frac{a+b}{RT} \right)\cdot
\sum_{\substack{s_1 \in ss[i+1,r-1],s_2 \in ss[r,j-1]\\ 
\text{$r$ base-paired in $s_2$}}} BF(s_1) \cdot BF(s_2) \cdot \left[
1+N(s_1)+N(s_2) \right]\\
&=& 
\sum_{r=i+5}^{j-5}
\exp \left( - \frac{a+b}{RT} \right)\cdot
\sum_{\substack{s_1 \in ss[i+1,r-1]\\ \text{~}}} BF(s_1)
\sum_{\substack{s_2 \in ss[r,j-1]\\ \text{$r$ base-paired in $s_2$}}} 
BF(s_2) + \\
&& 
\sum_{r=i+5}^{j-5}
\exp \left( - \frac{a+b}{RT} \right)\cdot
\sum_{\substack{s_1 \in ss[i+1,r-1]\\ \text{~}}} BF(s_1)N(s_1)
\sum_{\substack{s_2 \in ss[r,j-1]\\ \text{$r$ base-paired in $s_2$}}} 
BF(s_2) + \\
&&
\sum_{r=i+5}^{j-5}
\exp \left( - \frac{a+b}{RT} \right)\cdot
\sum_{\substack{s_1 \in ss[i+1,r-1]\\ \text{~}}} BF(s_1)
\sum_{\substack{s_2 \in ss[r,j-1]\\ \text{$r$ base-paired in $s_2$}}} 
BF(s_2) N(s_2)\\
&=& 
\exp \left( - \frac{a+b}{RT} \right)\cdot
\sum_{r=i+5}^{j-5}
\left[ ZM_{i+1,r-1} \cdot ZM1_{r,j-1} + \right. \\
&&
\left. QM_{i+1,r-1} \cdot ZM1_{r,j-1} + ZM_{i+1,r-1} \cdot QM1_{r,j-1} \right].
\end{eqnarray*}
Now $QB_{i,j} = A_{i,j}+B_{i,j}+C_{i,j}$. 
It nevertheless remains to define the recursions for $QM1_{i,j}$ and
$QM_{i,j}$.
These satisfy the following.
\begin{eqnarray*}
QM1_{i,j} &=&\sum_{k=i+4}^{j} 
\sum_{\substack{s \in ss[i,k]\\ \text{$(i,k) \in s$}}} 
\exp \left( - \frac{c(j-k)}{RT} \right)\cdot BF(s) \cdot \left[
N(s) + arc1(k+1,j) \right] \\
 &=&\sum_{k=i+4}^{j} 
\exp \left( - \frac{c(j-k)}{RT} \right)\cdot \left[
QB_{i,k} + ZB_{i,k} \cdot arc1(k+1,j) \right].
\end{eqnarray*}

\begin{eqnarray*}
QM_{i,j} &=&\sum_{r=i}^{j-5} 
%\exp \left( - \frac{b+c(r-i)}{RT} \right)\cdot QM1_{r,j} \cdot \left[
%arc1(i,r-1) + 1 \right] + \\
\exp \left( - \frac{b+c(r-i)}{RT} \right)\cdot \left[ QM1_{r,j} + ZM1_{r,j} 
\cdot arc1(i,r-1) \right] + \\
%arc1(i,r-1) + 1 \right] + \\
&& \sum_{r=i}^{j-5} 
\exp \left( - \frac{b}{RT} \right) \cdot \left[
QM_{i,r-1} ZM1_{r,j} +
ZM_{i,r-1} QM1_{r,j} \right].
\end{eqnarray*}
Suppose that $s = \{ (i,j),(i_1,j_1),\ldots,(i_k,j_k)\}$ 
is a multiloop closed by $(i,j)$, where $i<i_1<j_1<i_2<j_2 < \cdots
< i_k < j_k < j$. Then note that
we do not count neighbors of $s$ obtained by adding a base pair $(x,y)$
to the multiloop $s$, where $i< x < i_{\ell} < j_{\ell} < y$. 
Due to McCaskill's trick in the treatment of multiloops in the partition
function \cite{mcCaskill}, the treatment of such structural neighbors appears 
to be impossible while retaining the run time
$O(n^3)$. Nevertheless, multiloops are energetically costly due to
entropic considerations, and so penalized in the Turner energy model.
For this reason, multiloops are generally small, without many
unpaired bases $x,y$ capable of forming such base pairs. If a multiloop
is of sufficient size to permit such unpaired bases $x,y$, then the
multiloop free energy is likely to be large, 
so when the contribution is weighted by the Boltzmann factor of $s$, the
net contribution to $\langle N \rangle$ will be negligeable.

\hfill\break\newpage \clearpage
\begin{table*}
\begin{tabular}{|cccc|}
\hline
Rfam &Num Seq &EXPU& EXPB\\
\hline
RF00001 &712 & $0.367082 \pm 0.006445$ & $0.442810 \pm 0.057727$ \\
RF00004 &208 & $0.373414 \pm 0.004549$ & $0.611431 \pm 0.116160$ \\
RF00005 &960 & $0.373047 \pm 0.009376$ & $0.507313 \pm 0.122405$ \\
RF00008 &84  & $0.363520 \pm 0.012268$ & $0.413163 \pm 0.085666$ \\
RF00031 &61  & $0.369005\pm 0.008448$ & $0.572074 \pm 0.168601$ \\
RF00045 &66 & $0.371720 \pm 0.005207$ & $0.513964 \pm 0.068067$ \\
RF00167 &133 & $0.364369 \pm 0.009093$ & $0.731829 \pm 0.258182$ \\
RF00375 &130 & $0.356841 \pm 0.005719$ & $0.382784 \pm 0.058189$ \\
RF01055 & 160 & $0.368760 \pm 0.007053$ & $0.566054 \pm 0.124462$ \\
\hline
\end{tabular}
\caption{Normalized expected number of neighbors of sequences in the
seed alignment of various Rfam families \cite{Gardner.nar11}, given as the
mean plus or minus one standard deviation of values computed for each family.
Column 1 contains the Rfam family name of the noncoding RNA investigated.
Column 2 contains the number of sequences in the family; columns 3 and 4
respectively contain the average, taken over each Rfam family, of the
expected number of neighbors of each sequence, {\em normalized} by dividing
by sequence length, denoted respectively EXPU and EXPB. 
Computations were performed using
{\tt RNAexpNumNbors} with respect to the {\em uniform} probability 
(EXPU in column 3)
and {\em Boltzmann} probability (EXPB in column 4). The table clearly
shows that the Boltzmann expected number of neighbors is generally
larger than the uniform expected number of neighbors.
}
\label{table:RfamExpNumNborsNineFamilies}
\end{table*}

\begin{table}
\begin{tabular}{|l|cccc|}
\hline
& CMFE &  EXPB &  EXPU &   MFE \\
\hline
CMFE &   1.000000&       0.384740&       0.105773&       0.327990\\
EXPB &   0.384740&       1.000000&       0.028191&       0.777432\\
EXPU &   0.105773&       0.028191&       1.000000&       -0.003397\\
MFE  &   0.327990&       0.777432&       -0.003397&      1.000000\\
\hline
\end{tabular}
\caption{Pearson correlation between the length-normalized
number of neighbors of
the MFE structure, CMFE structure and the uniform or Boltzmann expected
value. Correlations were computed over the pooled sequences from all
nine Rfam families investigated.  Headers are explained as follows.
MFE: normalized number of neighbors for the MFE structure;
CMFE: normalized number of neighbors for the CMFE structure, i.e. the
structure having minimum free energy that is consistent with the Rfam
consensus structure;
EXPB: normalized expected number of neighbors with respect to 
Boltzmann probability;
EXPU: normalized expected number of neighbors with respect to 
uniform probability. Note the complete absence of correlation between
EXPB and EXPU. Results are similar when analyzing Rfam family
RF00001 (5S rRNA), both normalized and unnormalized (data not shown);
i.e. the lack of correlation is not due to any pooling effect of different
families. Although
Table~\ref{table:RfamExpNumNborsNineFamilies} shows that the Boltzmann
expected number of neighbors is greater than the uniform expected number
of neighbors, there appears to be no correlation between these values.
}
\label{table:normNumNbors}
\end{table}

\begin{table}
\begin{tabular}{|l|ccccc|}
\hline
& H &  EnsDef&          ExpBPdist&  ExpNumBP&  ExpNumNbors\\
\hline
H& 1.000000& 0.716498&        0.699402&        -0.471001&       -0.000638\\
EnsDef& 0.716498& 1.000000&   0.997092&        -0.306858&       -0.024428\\
ExpBPdist &0.699402&  0.997092& 1.000000&      -0.285362&       -0.027360\\
ExpNumBP& -0.471001& -0.306858 & -0.285362&     1.000000&       -0.006972\\
ExpNumNbors& -0.000638 &-0.024428& -0.027360&   -0.006972&       1.000000\\
\hline
\end{tabular}
\caption{Correlations of various measures that depend on the Boltzmann
ensemble of all secondary structures of a given RNA sequence. All
measures have been normalized by dividing by sequence length. Column and
row headers are explained as follows.
H: average positional entropy,
EnsDef: ensemble defect to MFE structure,
ExpBPdist: expected base pair distance to MFE structure,
ExpNumBP: expected number of base pairs,
ExpNumNbors: expected number of neighbors.
See Appendix for definitions of these measures.
}
\label{table:noCorrWithOtherMeasures}.
\end{table}

\begin{table}
\begin{tabular}{|l|cccccc|}
\hline
  & E &MFE-EXPB &EXPB &MFE-EXPU & MFE &SeqLen \\
\hline
E &1.000000 &0.088782 &0.138171 &0.135036 &0.132890 &-0.862103 \\
MFE-EXPB &0.088782 &1.000000 &0.370294 &0.872449 &0.872135 &-0.000543\\ 
EXPB &0.138171 &0.370294 &1.000000 &0.775958 &0.777432 &0.125386 \\
MFE-EXPU &0.135036 &0.872449 &0.775958 &1.000000 &0.999436 &0.065343 \\
MFE &0.132890 &0.872135 &0.777432 &0.999436 &1.000000 &0.065674 \\
SeqLen &-0.862103 &-0.000543 &0.125386 &0.065343 &0.065674 &1.000000 \\ 
%A &1.000000 &0.088782 &0.138171 &0.135036 &0.132890 &-0.862103 \\
%B &0.088782 &1.000000 &0.370294 &0.872449 &0.872135 &-0.000543\\ 
%C &0.138171 &0.370294 &1.000000 &0.775958 &0.777432 &0.125386 \\
%D &0.135036 &0.872449 &0.775958 &1.000000 &0.999436 &0.065343 \\
%E &0.132890 &0.872135 &0.777432 &0.999436 &1.000000 &0.065674 \\
%F &-0.862103 &-0.000543 &0.125386 &0.065343 &0.065674 &1.000000 \\ 

\hline
\end{tabular}
\caption{Correlations for data pooled from all nine Rfam families.
Column headers designate the following.
E: minimum free energy;
MFE-EXPB: (number of neighbors of MFE structure minus the Boltzmann expected number of neighbors)
divided by sequence length -- i.e. length-normalized;
EXPB: Boltzmann expected number of neighbors, divided by sequence length;
MFE-EXPU: (number of neighbors of MFE structure minus the uniform expected number of neighbors)
divided by sequence length;
MFE: number of neighbors of the MFE structure divided by sequence length;
SeqLen: sequence length. 
}
\label{table:corrNormDiffA}
\end{table}

\begin{table}
\begin{tabular}{|l|cccc|}
\hline 
FASTA&	 MFE-EXP&	 CMFE-MFE&	 CMFE-EXP&	BPdist\\
\hline 
RF00001&	$0.0293 \pm 0.0881$&	$0.2006 \pm 0.5482$&	$0.2300 \pm 0.5527$&	$0.1884 \pm 0.1614$\\
RF00004&	$0.0543 \pm 0.1872$&	$0.2863 \pm 0.6817$&	$0.3406 \pm 0.6945$&	$0.2082 \pm 0.1243$\\
RF00005&	$0.0447 \pm 0.1863$&	$0.4260 \pm 0.6458$&	$0.4707 \pm 0.6195$&	$0.2631 \pm 0.1265$\\
RF00008&	$0.1336 \pm 0.0998$&	$0.0133 \pm 0.0665$&	$0.1469 \pm 0.1024$&	$0.0492 \pm 0.1143$\\
RF00031&	$0.0772 \pm 0.3079$&	$0.1895 \pm 0.4428$&	$0.2667 \pm 0.5247$&	$0.1108 \pm 0.1207$\\
RF00045&	$0.0364 \pm 0.1028$&	$0.5857 \pm 1.0071$&	$0.6221 \pm 0.9906$&	$0.3064 \pm 0.1882$\\
RF00167&	$0.2412 \pm 0.3268$&	$0.0714 \pm 0.3087$&	$0.3126 \pm 0.4151$&	$0.0631 \pm 0.1175$\\
RF00375&	$0.0344 \pm 0.1122$&	$0.0012 \pm 0.1524$&	$0.0356 \pm 0.1711$&	$0.1018 \pm 0.1455$\\
RF01055&	$0.1148 \pm 0.1812$&	$0.2753 \pm 0.3010$&	$0.3901 \pm 0.2401$&	$0.2042 \pm 0.1400$\\
\hline 
\end{tabular}
\caption{Table to determine whether the minimum free energy structure
has more neighbors than the expected number. MFE [resp. CMFE] stands for
the {\em length-normalized} number of neighbors for the minimum free
energy structure [resp. the structure having minimum free energy among
those structures that are consistent with the Rfam consensus structure].
EXP stands for the {\em length-normalized} expected number of neighbors,
as computed by {\tt RNAexpNumNbors}, and BPdist is the {\em length-normalized}
base pair distance between the MFE structure and the CMFE structure.
Values are given as the mean
plus or minus one standard deviation, taken over all sequences in the
corresponding Rfam family.
}
\label{table:expLessMfeLessCmfe}
\end{table}

\bibliographystyle{plain}
%\bibliography{/Users/clote/text/BIBdir/clote}
%\bibliography{biblio}

\end{document}